\documentclass[aps,prd,twocolumn,superscriptaddress,showpacs,floatfix]{revtex4-1}

\usepackage{amsmath,amssymb}
\usepackage{graphicx}% Include figure files
\usepackage{dcolumn}% Align table columns on decimal point
\usepackage{bm}% bold math
\usepackage{color}
\usepackage{comment}

\newcommand{\ASCm}{ASC$_-$}
\newcommand{\ASCp}{ASC$_+$}
\newcommand{\ASCpm}{ASC$_{\pm}$}
\newcommand{\acot}{\text{acot\,}}
\newcommand{\atan}{\text{atan\,}}
\newcommand{\atanh}{\text{atanh\,}}
\newcommand{\band}{\text{band}}
\newcommand{\bk}{\mathbf{k}}
\newcommand{\dn}{\downarrow}
\newcommand{\ds}{\displaystyle}

\newcommand{\Ed}{\veps_d}
\newcommand{\Ek}{\veps_{\bk}}
\newcommand{\G}{\Gamma}
\newcommand{\Gam}{\bar{\G}}
\renewcommand{\H}{\hat{H}}
\newcommand{\half}{{\textstyle\frac{1}{2}}}
\newcommand{\imp}{\text{imp}}
\newcommand{\hyb}{\text{hyb}}
\newcommand{\LM}{\text{LM}}
\newcommand{\MV}{\text{MV}}
\newcommand{\n}{\hat{n}}
\newcommand{\p}{\perp}
\newcommand{\pdag}{\phantom{\dag}}
\newcommand{\ph}{\textit{p-h}}
\newcommand{\ptD}{\vphantom{\tD}D}
\newcommand{\s}{\sigma}

\newcommand{\tD}{\tilde{D}}
\newcommand{\tE}{\tilde{E}}
\newcommand{\tEd}{\tilde{\veps}_d}
\newcommand{\tEdone}{\tilde{\veps}_{d,1}}
\newcommand{\tG}{\tilde{\G}}
\newcommand{\tj}{\tilde{\jmath}}
\newcommand{\tjp}{\tj_{\!\p}}
\newcommand{\tJ}{\tilde{J}}
\newcommand{\ts}{\textstyle}
\newcommand{\tU}{\tilde{U}}
\newcommand{\tV}{\tilde{V}}
\newcommand{\tx}{\tilde{x}}
\newcommand{\up}{\uparrow}
\newcommand{\veps}{\varepsilon}

\begin{document}

\preprint{APS/123-QED}

\title{Phase boundaries of power-law Anderson and Kondo models:
A poor man's scaling study}

\author{Mengxing Cheng}
\affiliation{Department of Physics, University of Florida,
P.O.\ Box 118440, Gainesville, Florida 32611-8440, USA}
\affiliation{Research Computing Center, University of Chicago,
5607 South Drexel Avenue, Chicago, Illinois 60637, USA}

\author{Tathagata Chowdhury}
\email{tatha@thp.uni-koeln.de}
\affiliation{Department of Physics, University of Florida,
P.O.\ Box 118440, Gainesville, Florida 32611-8440, USA}
\affiliation{Institut f\"{u}r Theoretische Physik, Universit\"{a}t zu K\"{o}ln,
Z\"{u}lpicher Stasse 77a, 507937 K\"{o}ln, Germany}

\author{Aaron Mohammed}
\affiliation{Department of Physics, University of Florida,
P.O.\ Box 118440, Gainesville, Florida 32611-8440, USA}
\affiliation{Department of Physics, University of South Florida,
4202 East Fowler Avenue, Tampa, Florida 33620, USA}

\author{Kevin Ingersent}
\affiliation{Department of Physics, University of Florida,
P.O.\ Box 118440, Gainesville, Florida 32611-8440, USA}

\date{\today}

\begin{abstract}
We use the poor man's scaling approach to study the phase boundaries of a
pair of quantum impurity models featuring a power-law density of states
$\rho(\veps)\propto|\veps|^r$, either vanishing (for $r>0$) or
diverging (for $r<0$) at the Fermi energy $\veps=0$, that gives rise to
quantum phase transitions between local-moment and Kondo-screened phases.
For the Anderson model with a pseudogap (i.e., $r>0$), we find the phase
boundary for (a) $0<r<1/2$, a range over which the model exhibits
interacting quantum critical points both at and away from particle-hole
(\ph) symmetry, and (b) $r>1$, where the phases are separated by
first-order quantum phase transitions that are accessible only for broken
\ph\ symmetry.
For the \ph-symmetric Kondo model with easy-axis or easy-plane anisotropy
of the impurity-band spin exchange, the phase boundary and scaling
trajectories are obtained for both $r>0$ and $r<0$.
Throughout the regime of weak-to-moderate impurity-band coupling in
which poor man's scaling is expected to be valid, the approach predicts
phase boundaries in excellent qualitative and good quantitative agreement
with the nonperturbative numerical renormalization group, while also
establishing the functional relations between model parameters along
these boundaries.
\end{abstract}


\maketitle

\section{Introduction}
\label{sec:intro}

The Kondo problem---the question of how an impurity local moment becomes
screened at low temperatures by the conduction electrons of a host
metal---has been highly influential in stimulating the development of
theoretical and numerical methods for treating strongly correlated condensed
matter \cite{Hewson:93}. Perturbative treatments of the spin-flip scattering
between local and delocalized spins necessarily break down below a
characteristic Kondo temperature scale, giving rise to a complex many-body
problem. Nonetheless, much valuable understanding of the Kondo problem has
come from perturbative renormalization-group (RG) \cite{Abrikosov:70,Fowler:71}
and perturbative scaling \cite{Anderson:70a,Anderson:70b} approaches. These
were distilled into their simplest form in the poor man's scaling of Anderson
\cite{Anderson:70b}.

In poor man's scaling, electron states far from the Fermi energy are
progressively eliminated as the effective bandwidth is reduced with a
compensating adjustment of a dimensionless measure of the effective
impurity-band exchange coupling. The evolution of this coupling to ever larger
values with decreasing bandwidth is suggestive of approach to a fully screened
strong-coupling fixed point, although the scaling approach breaks down once
the effective bandwidth drops below the order of the Kondo temperature. More
sophisticated but generally less intuitive methods (the first historically
being the numerical renormalization group or NRG \cite{Wilson:75}) were
devised to confirm that the infrared fixed point indeed corresponds to
infinite exchange \cite{Hewson:93}. Poor man's scaling was subsequently
extended to the Anderson model with impurity Coulomb interaction $U=\infty$
\cite{Jefferson:77,Haldane:78} and the $n$-channel Kondo model
\cite{Nozieres:80}, where it correctly predicts the existence of a stable RG
fixed point at an intermediate value of the exchange coupling that lies within
the perturbative domain for $n\gg 2$.

More recently, there has been much interest in Kondo physics in settings where
the band density of states has a power-law variation
$\rho(\veps)\propto|\veps|^r$ in the vicinity of the Fermi energy $\veps=0$.
Pseudogaps described by exponents $r>0$ can be found in a variety of materials
including heavy-fermion and cuprate unconventional superconductors
\cite{Sigrist:91,Timusk:99}, zero-gap bulk \cite{Hohler:83} and engineered
\cite{Volkov:85} semiconductors, and various (quasi-)two-dimensional systems
such as graphite \cite{DiVincenzo:84,Semenoff:84} and graphene
\cite{CastroNeto:09}. An exponent $r=-\half$ arises near a band edge in
one-dimensional leads, while values $-1<r<0$ can describe disordered Dirac
fermions in two dimensions \cite{Ludwig:94, Motrunich:02}.
Several theoretical techniques that have proved powerful for describing quantum
impurities in metallic hosts, including the Bethe ansatz, bosonization, and
conformal field theory, cannot be applied for a power-law density of states.
However, power-law variants of the Kondo impurity model and the corresponding
Anderson model have been extensively studied using other techniques such as
perturbative scaling \cite{Withoff:90,Ingersent:96,Gonzalez-Buxton:96,Vojta:02,
Mitchell:13}, large-$N$ approaches \cite{Withoff:90,Borkowski:92,Cassanello:96,
Vojta:01a,Polkovnikov:02}, the NRG \cite{Chen:95,Ingersent:96a,Bulla:97,
Gonzalez-Buxton:98,Vojta:01b,Vojta:02,Ingersent:02,Pixley:12,Chowdhury:15},
the perturbative RG \cite{Kircan:04,Vojta:04,Fritz:04}, and the local-moment
approach \cite{Logan:00,Bulla:00,Glossop:03}. Due to the depletion of the
conduction-band density of states near the Fermi energy, these pseudogap models
feature quantum phase transitions \cite{Withoff:90} between a local-moment
phase for weak impurity-band couplings, in which the impurity spin survives
unscreened down to zero temperature, and one or more strong-coupling Kondo
phases in which the impurity spin undergoes complete or partial many-body
screening (depending on the presence or absence of particle-hole symmetry)
\cite{Gonzalez-Buxton:98}.

Of all the techniques so far applied to the pseudogap Anderson and Kondo
models, only the NRG has proved capable of capturing all the key features of
the phase diagram, including the existence of four qualitatively different
ranges of the band exponent $r$ \cite{Gonzalez-Buxton:98}. However, as is true
for many computational methods, the NRG's reliability comes at the price of
laborious implementation and a loss of physical transparency.
Together, these make it difficult to obtain simple intuition about how two
fundamentally opposing tendencies---growth of host correlations engendered
by a local dynamical degree of freedom, and the weakening of host-impurity
interaction due to depression of the low-energy density of states---compete
to create nontrivial temperature dependencies of physical properties and to
shape phase boundaries. The local-moment approach \cite{Logan:98} reproduces
rather well the phase boundaries of the pseudogap Anderson model with band
exponents $0 < r < 1$, but its analytical insights are confined to situations
of strict particle-hole symmetry \cite{Logan:00, Bulla:00} or the limit
$r\to 0^+$ \cite{Glossop:03}.

It is highly desirable to identify another primarily analytical approach that
can shed light more widely on the functional relations describing the phase
boundaries in challenging quantum impurity problems that feature both
(i) more than one independent coupling that flows under the reduction of the
effective bandwidth, and (ii) unstable quantum critical points arising from
competing flows in the multidimensional parameter space of effective couplings.
A promising candidate is poor man's scaling \cite{Anderson:70b}, which has
previously been established to account well for the possible ground states of
many quantum impurity problems and to provide an approximate description of
the physics on different energy/temperature scales in terms of a flow through
a space of renormalized Hamiltonian couplings. The method yields a set of
ordinary differential equations describing the renormalization of Hamiltonian
couplings. These differential equations can in some cases be integrated in
closed form; failing that, their solutions can be explored numerically via
numerical iteration from different choices of bare couplings.

In this paper, we critically evaluate the adequacy of poor man's scaling
for describing phase boundaries in the Anderson model (with an arbitrary
on-site repulsion $U$) and in the particle-hole-symmetric Kondo model with
easy-axis or easy-plane anisotropy of the impurity-band exchange coupling.
For each model, we generalize previous treatments to obtain coupled
differential equations for the evolution of effective couplings under
progressive reduction of the conduction bandwidth. These equations are valid
for any density of states of the form $\rho(\omega)\propto|\omega|^r$, whether
$r$ is positive, negative, or zero. (The case $r=0$ describes conventional
metallic hosts.)
We obtain analytical expressions for the locations of phase boundaries for
different parameter ranges of the pseudogap ($r>0$) Anderson and power-law
($r\ne 0$) anisotropic Kondo models. Comparison with nonperturbative NRG
results shows that throughout the perturbative regime where the method is
well-founded, poor man's scaling correctly captures the functional
relations between model parameters along various parts of the phase
boundaries, and also reproduces the absolute location of the boundaries
with good quantitative accuracy. The availability of proven analytical
expressions obviates the need for further NRG calculations to understand
and make predictions about possible realizations of these models.

The rest of the paper is organized as follows. Section \ref{sec:Anderson}
treats the Anderson model with a power-law density of states. Section
\ref{subsec:model:A} defines the model and summarizes the phase diagram that has
been established through previous work. The poor man's scaling equations are
derived in Sec.\ \ref{subsec:poor man:A}. Section \ref{subsec:r>1} compares
analytic approximations for the phase boundary with NRG results for superlinear
($r>1$) densities of states and various ranges of the other model parameters,
while Sec.\ \ref{subsec:r<1} does the same for $0 < r < 1$. The anisotropic
Kondo model is the subject of Sec. \ref{sec:Kondo}.
Section \ref{subsec:scaling:K} presents the poor man's scaling equations along
with a preliminary analysis. Phase boundaries are analyzed for $0<r<\half$ and
$-1<r<0$ in Secs.\ \ref{subsec:pseudogap:K} and \ref{subsec:divergent:K},
respectively. Section \ref{sec:discussion} contains a brief discussion of
strengths and weaknesses shown by the poor man's scaling approach.

\section{Power-Law Anderson model}
\label{sec:Anderson}

\subsection{Model Hamiltonian}
\label{subsec:model:A}
The Anderson impurity model is described by the Hamiltonian \cite{Anderson:61}
\begin{equation}
\label{H_A}
\H_A = \H_{\band} + \H_{\imp} + \H_{\hyb} \, ,
\end{equation}
where
\begin{equation}
\label{H_band}
\H_{\band}
= \sum_{\bk,\s}\veps^{\pdag}_{\bk}c^{\dag}_{\bk\s}
  c^{\pdag}_{\bk \s}
\end{equation}
with $\sigma=\pm 1$ (or $\sigma=\:\up$, $\dn$) describes a noninteracting
conduction band having dispersion $\Ek$;
\begin{equation}
\label{H_imp}
\H_{\imp} = \Ed \, \n_d + U \n_{d\up} \n_{d\dn}
\end{equation}
with $\n_d=\n_{d\up}+\n_{d\dn}$ and $\n_{d\s}=d_{\s}^{\dag} d_{\s}^{\pdag}$
describes an impurity having level energy $\Ed$ and on-site
Coulomb interaction $U$; and the hybridization term
\begin{equation}
\label{H_hyb}
\tilde{H}_{\hyb} = \frac{1}{\sqrt{N_k}} \sum_{\bk,\s}
   \bigl( V_{\bk} d_{\s}^{\dag} c^{\pdag}_{\bk\s} + \text{H.c.} \bigr)
\end{equation}
accounts for impurity-band tunneling. $N_k$ is the number of unit cells in the
host metal, i.e., the number of inequivalent $\bk$ values. Without loss of
generality, we take the hybridization matrix element $V_{\bk}$ to be real and
non-negative. For compactness of notation, we drop all factors of the reduced
Planck constant $\hbar$, Boltzmann's constant $k_B$, and the impurity magnetic
moment $g\mu_B$.

A mapping to an energy representation where
\begin{align}
\H_{\band}
&= \sum_{\s} \int\!d\veps \: \veps \, c^{\dag}_{\veps\s}
   c^{\pdag}_{\veps\s} , \\
\H_{\hyb}
&= \sum_{\s} \int\!d\veps \: \sqrt{\Gam(\veps)/\pi} \:
   \bigl( \veps c^{\dag}_{\veps\s} d^{\pdag}_{\s} + \text{H.c.} \bigr) ,
\end{align}
shows that the conduction-band dispersion $\veps_{\bk}$ and the hybridization
matrix element $V_{\bk}$ affect the impurity degrees of freedom only in
combination through the hybridization function \cite{Gamma-notation}
\begin{equation}
\label{Gamma:def}
\Gam(\veps) \equiv \frac{\pi}{N_k} \sum_{\bk}
   V_{\bk}^2 \, \delta(\veps-\veps_{\bk}) .
\end{equation}
To focus on the most interesting physics of the model, we assume a simplified
form
\begin{equation}
\label{Gamma:power}
\Gam(\veps) = \G \, |\veps/D|^r \, \Theta(D-|\veps|) ,
\end{equation}
where $\Theta(x)$ is the Heaviside function and $\Gamma$ is the hybridization
width. The primary focus of this work is cases $r>0$ in which the hybridization
function exhibits a power-law pseudogap around the Fermi energy. We will also
briefly discuss $r=0$, representing a conventional metallic host.

One way that a hybridization function of the form of Eq.\ \eqref{Gamma:power}
can arise is from a purely local hybridization matrix element $V_{\bk}=V\ge 0$
combined with a density of states (per unit cell, per spin orientation) varying
as
\begin{equation}
\label{rho:def}
\rho(\veps) \equiv N_k^{-1} \sum_{\bk} \, \delta(\veps-\veps_{\bk})
= \rho_0|\veps/D|^r\Theta(D-|\veps|)
\end{equation}
with $\rho_0=(1+r)/(2D)$, in which case $\G=\pi\rho_0 V^2$.
However, all results below apply equally to situations in which the $\bk$
dependence of the hybridization contributes to the energy dependence of
$\Gam(\veps)$.

The assumption that $\Gam(\veps)$ exhibits a pure power-law dependence
over the entire width of the conduction band is a convenient idealization. More
realistic hybridization functions in which the power-law variation is restricted
to a region around the Fermi energy exhibit the same qualitative physics, with
modification only of nonuniversal properties such as the location of
phase boundaries and the value of the Kondo temperature.

\begin{figure}[t]
\centering
\includegraphics[width=.95\columnwidth]{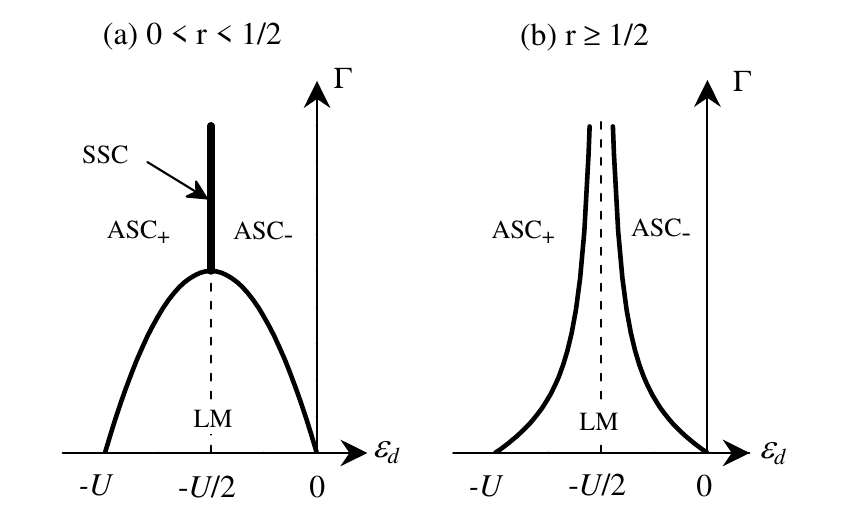}
\caption{\label{fig:PAM_phase}
Schematic $\Ed$-$\G$ phase diagrams of the pseudogap Anderson model
[Eqs.\ \protect\eqref{H_A}--\protect\eqref{Gamma:power}]
for band exponents (a) $0 < r < \half$, (b) $r\ge\half$. Generically, the
system falls into either a local-moment phase (LM) or one of two asymmetric
strong-coupling phases (\ASCpm). However, there is also a symmetric
strong-coupling phase (the line labeled SSC) that is reached only for
$0<r<\half$ under conditions of strict particle-hole symmetry
($\Ed=-\half U$) and for sufficiently large hybridization widths $\G$.}
\end{figure}

In the metallic ($r=0$) Anderson model, any value $\G>0$ places the system
in its strong-coupling phase, where the impurity degrees of freedom are
completely quenched at $T=0$. The situation for
pseudogapped hybridization functions ($r>0$) is much richer, as summarized in
the phase diagrams shown in Fig.\ \ref{fig:PAM_phase} for cases $U>0$ of on-site
Coulomb repulsion. The most notable feature is the existence within a region
$-U<\Ed<0$, $\G<\G_c(r,U,\Ed)$ of a local-moment (LM) phase within which the
impurity retains an unquenched spin degree of freedom down to $T=0$. There
are also three different strong-coupling phases, distinguished by their
ground-state electron number $Q$ (measured from half filling): a symmetric
strong-coupling (SSC) phase with $Q=0$, reached only for $0<r<\half$ under
the condition $\Ed=-\half U$ for strict particle-hole (\ph) symmetry;
and a pair of asymmetric strong-coupling phases \ASCp\ and \ASCm\ having
$Q=1$ and $Q=-1$, respectively. The ranges $0<r\le \half$ and $r\ge \half$
can both be further subdivided based on the nature of the quantum phase
transitions separating the phases. For a compact summary, the reader is
referred to Sec.\ II\ B1 of Ref.\ \onlinecite{Cheng:13}.

\subsection{Derivation of poor man's equations}
\label{subsec:poor man:A}

This section presents a poor man's scaling treatment of the Anderson
Hamiltonian with a power-law hybridization function. Jefferson
\cite{Jefferson:77} and Haldane \cite{Haldane:78} provided scaling treatments
of the metallic case $r=0$ valid in the limit $U\gg D$. These were subsequently
extended to general values of $U$ (Ref.\ \onlinecite{Hewson:93}), although the
analysis neglected the renormalization of $U$. Reference
\onlinecite{Gonzalez-Buxton:96} presented scaling equations for the pseudogap
case $r>0$ with $U=\infty$. Below, the scaling analysis is generalized to
arbitrary values of $r$ and $U$. Two of us have previously presented a similar
poor man's scaling analysis of the Anderson-Holstein impurity model with a
power-law hybridization \cite{Cheng:13}. The treatment of the Anderson model
is somewhat simpler, and as we will see, the resulting scaling equations are
amenable to approximate integration in several physically interesting limits.

We start with the Anderson Hamiltonian written in the form
\begin{equation}
\H'_A = \H_{\band} + \H_{\imp} + \H'_{\hyb} \, ,
\end{equation}
where $\H_{\band}$ and $\H_{\imp}$ are as defined in Eqs.\ \eqref{H_band} and
\eqref{H_imp}, respectively, but with $\H_{\hyb}$ in Eq.\ \eqref{H_hyb}
replaced by
\begin{align}
\label{H_hyb2}
\hat{H}'_{\hyb}
&= \frac{1}{\sqrt{N_k}} \sum_{\bk,\s}
   \bigl\{ \bigl[ V_{0,\bk} (1-\n_{d,-\s}) \notag \\
&\qquad + V_{1,\bk} \, \n_{d,-\s} \bigr] d^{\dag}_{\s}
  c^{\pdag}_{\bk\s} + \text{H.c.} \bigr\},
\end{align}
with hybridization functions
\begin{equation}
\label{Gamma_tau:def}
\Gam_{\tau}(\veps)=\frac{\pi}{N_k}\sum_{\bk} V_{\tau,\bk}^2 \,
  \delta(\veps-\Ek) = \G_{\tau}|\veps/D|^r \: \Theta(D-|\veps|)
\end{equation}
for $\tau = 0$, $1$ having the same power-law dependence as $\Gam(\veps)$
defined in Eq.\ \eqref{Gamma:power}. At the bare Hamiltonian level, one
expects the hybridization $V_{0,\bk}$ between the empty and singly occupied
impurity configurations to be identical to the matrix element $V_{1,\bk}$
between the singly occupied and doubly occupied impurity configurations.
However, this degeneracy might be broken under the scaling procedure.

Following Haldane \cite{Haldane:78}, we focus on many-body states $|0\rangle$,
$|\s\rangle=d_{\s}^{\dag}|0\rangle$, and
$|2\rangle=d_{\up}^{\dag}d_{\dn}^{\dag}|0\rangle$ formed by combining the
conduction-band ground state (having $N_k$ electrons of energy $\Ek<0$)
with one of the four possible configurations of the impurity level. Neglecting
for the moment the effect of the hybridization ($\H'_{\hyb}$), the energies of
these states are denoted $E_0$, $E_1=E_0+\Ed$, and $E_2=E_1+\Ed+U=2E_1-E_0+U$.

We now consider the effect of an infinitesimal reduction in the half-bandwidth
from $D$ to $\tD=D+dD$, where $dD < 0$. The goal is to write a new
Hamiltonian $\tilde{H}'_A$ similar in form to $\H'_A$ but retaining only
conduction-band degrees of freedom having energies $|\Ek|<\tD$ and
with parameters $\tEd$, $\tU$, and $\tV_{\tau,\bk}$ adjusted to account
perturbatively for the band-edge states that have been eliminated.

Let $K^+$ be the set of wave vectors $\bk$ describing particle-like states
having energies $\tD<\Ek<D$, and $K^-$ be the set of wave vectors describing
hole-like state with energies $-D<\Ek<-\tD$. Virtual tunneling of an electron
from a $K^-$ state into the empty impurity level transforms the state
$|0\rangle$ to
\begin{equation}
\label{tilde0}
|\tilde{0}\rangle = |0\rangle + \sum_{\s}\frac{1}{\sqrt{N_k}}\sum_{\bk\in K^-}
  \frac{V_{0,\bk}}{|\Ek|+E_1-E_0} \; c_{\bk\s} |\s\rangle + O(V^2)
\end{equation}
with energy
\begin{align}
\label{tildeE_0}
\tE_0
&= E_0 - \frac{2}{N_k} \sum_{\bk\in K^-}
   \frac{V_{0,\bk}^2}{|\Ek|+E_1-E_0} + O(V^3) \notag \\
&\simeq E_0 - \frac{|dD|}{\pi} \, \frac{2\Gam_0(-D)}{D+\Ed} + O(V^3) \, .
\end{align}
Here, $O(V^n)$ schematically represents all processes involving a product
of least $n$ factors $V_{\tau_j,\bk_j}$.
Similarly, virtual tunneling of an electron from the doubly occupied impurity
level into a $K^+$ state transforms $|2\rangle$ to
\begin{equation}
\label{tilde2}
|\tilde{2}\rangle = |2\rangle + \sum_{\s}\frac{\s}{\sqrt{N_k}}\sum_{\bk\in K^+}
  \frac{V_{1,\bk}}{\Ek+E_1-E_2} \; c^{\dag}_{\bk\s}
  |\!-\!\s\rangle + O(V^2)
\end{equation}
with energy
\begin{align}
\label{tildeE_2}
\tE_2
&= E_2 - \frac{2}{N_k} \sum_{\bk\in K^+}
  \frac{V_{1,\bk}^2}{\Ek+E_1-E_2} + O(V^3) \notag \\
&\simeq E_2 - \frac{|dD|}{\pi} \, \frac{2\Gam_1(D)}{D-U-\Ed} + O(V^3) \, .
\end{align}
Finally, virtual tunneling of an electron into the singly occupied impurity
from a $K^-$ state or from the singly occupied level into a $K^+$ state
transforms $|\s\rangle$ to
\begin{align}
\label{tilde1}
|\tilde{\s}\rangle
&= |\s\rangle - \frac{\sigma}{\sqrt{N_k}} \sum_{\bk\in K^-}
  \frac{V_{1,\bk}}{|\Ek|+E_2-E_1} \; c_{\bk,-\s} |2\rangle
  \notag \\
& \quad - \frac{1}{\sqrt{N_k}} \sum_{\bk\in K^+}
    \frac{V_{0,\bk}}{\Ek+E_0-E_1} \; c^{\dag}_{\bk\s} |0\rangle + O(V^2)
\end{align}
with energy
\begin{align}
\label{tildeE_1}
\tE_1
&= E_1 - \frac{1}{N_k} \sum_{\bk\in K^-}
   \frac{V_{1,\bk}^2}{|\Ek|+E_2-E_1} \notag \\
& \qquad - \frac{1}{N_k} \sum_{\bk\in K^+}
   \frac{V_{0,\bk}^2}{\Ek+E_0-E_1} + O(V^3) \notag \\
&\simeq E_1 - \frac{|dD|}{\pi} \left[ \frac{\Gam_1(-D)}{D+U+\Ed}
   + \frac{\Gam_0(D)}{D-\Ed} \right] + O(V^3).
\end{align}
The $O(V^2)$ terms in each of the above states $|\tilde{\phi}\rangle$ include
ones required to enforce normalization, i.e.,
$\langle\tilde{\phi}|\tilde{\phi}\rangle = \langle\phi|\phi\rangle = 1$.

The modified energies can be used to define effective Hamiltonian parameters
$\tEd=\tE_1-\tE_0$ and $\tU=\tE_2+\tE_0-2\tE_1$. At the same time, for each
$\bk$ in the retained portion of the band (i.e., satisfying $|\Ek|<\tD$),
the hybridization matrix element $V_{0,\bk}$ must be replaced by
\begin{equation}
\tV_{0,\bk} =
  \begin{cases}
    \sqrt{N_k} \: \langle \tilde{0}|c_{\bk\s} \H'_A|\tilde{\s}\rangle
    & \text{for\;\;} \Ek>0 \\[1ex]
    - \sqrt{N_k} \: \langle \tilde{\s}| c_{\bk\s}^{\dag} \H'_A|\tilde{0}\rangle
    & \text{for\;\;} \Ek<0 ,
  \end{cases}
\end{equation}
and $V_{1,\bk}$ must be replaced by
\begin{equation}
\tV_{1,\bk} =
  \begin{cases}
    -\sigma \sqrt{N_k}
      \: \langle\tilde{\s}|c_{\bk,-\s} \H'_A|\tilde{2}\rangle
    & \text{for\;\;} \Ek>0 \\[1ex]
    \sigma\sqrt{N_k}
      \: \langle \tilde{2}| c_{\bk,-\s}^{\dag} \H'_A|\tilde{\s}\rangle
    & \text{for\;\;} \Ek<0.
  \end{cases}
\end{equation}
It is straightforward to show that
\begin{equation}
\label{tildeV_tau,k}
\tV_{\tau,\bk} = V_{\tau,\bk} + O(V^3).
\end{equation}
The leading corrections to $\tV_{\tau,\bk}$ involve numerous terms
arising from the $V^2$ terms in the states $|\tilde{\phi}\rangle$. Since these
corrections are too small to be of much practical importance, we shall not
evaluate them here.

The infinitesimal band-edge reduction described in the previous paragraphs
can be carried out repeatedly to reduce the half-bandwidth by a finite
amount from $D$ to $\tD<D$. Equations \eqref{tildeE_0} and \eqref{tildeE_1}
indicate that during this process, the effective impurity level energy
$\tEd=\tE_1-\tE_0$ evolves according to the scaling equation
\begin{equation}
\label{Ed:scaling1}
\frac{d\tEd}{d\tD}
  = \frac{1}{\pi} \biggl[ \frac{\tG_{0,+}}{\tD-\tEd}
  - \frac{2\tG_{0,-}}{\tD+\tEd}
  + \frac{\tG_{1,-}}{\tD+\tU+\tEd} \, \biggr] + O(V^3) ,
\end{equation}
where $\tG_{\tau,\pm}$ is the rescaled hybridization function at the reduced
band edges $\veps=\pm\tD$. Taking into account Eq.\ \eqref{tildeE_2} as well,
one sees that the effective on-site repulsion $\tU=\tE_2+\tE_0-2\tE_1$ follows
\begin{align}
\label{U:scaling1}
\frac{d\tU}{d\tD}
&= \frac{2}{\pi} \biggl[ \frac{\tG_{0,-}}{\tD+\tEd}
   - \frac{\tG_{0,+}}{\tD-\tEd} + \frac{\tG_{1,+}}{\tD-\tU-\tEd}
   \notag \\
&\qquad \quad - \frac{\tG_{1,-}}{\tD+\tU+\tEd} \, \biggr] + O(V^3) .
\end{align}
The band-edge values $\tG_{\tau,\pm}$ of the hybridization functions
$\Gam_{\tau}(\veps)$ rescale both due to
the replacement of $D$ by $\tD$ in Eq.\ \eqref{Gamma:power} and due to the
perturbative corrections to $V_{\tau,\bk}$ in Eq.\ \eqref{tildeV_tau,k},
leading to the scaling equation
\begin{equation}
\label{Gamma_tau:scaling}
\frac{d\tG_{\tau,\pm}}{d\tD} = r\,\frac{\tG_{\tau,\pm}}{\tD} + O(V^4).
\end{equation}
The scaling equations \eqref{Ed:scaling1}--\eqref{Gamma_tau:scaling} have been
derived to lowest order in nondegenerate perturbation theory, and are strictly
valid only so long as $|\tD\pm(\tEd+\tau U)| \gg \tV_{\tau,\bk}$ for each $\bk$
such that $\veps_{\bk}=\mp\tD$.

Equation \eqref{Gamma_tau:scaling} shows that the band-edge values of the
hybridization functions $\Gam_{\tau}(\veps)$ are irrelevant (in the RG sense)
for $r>0$ and at most marginally relevant
%or marginally irrelevant
for $r=0$.
For the \ph-symmetric bare hybridization functions considered in this
work, it is an excellent approximation to set $\tG_{0,\pm}=\tG_{1,\pm}=\tG$,
leading to the simplified scaling equations
\begin{align}
\label{Gamma:scaling}
\frac{d\tG}{d\tD}
&= r\,\frac{\tG}{\tD} \, , \\
\label{Ed:scaling}
\frac{d\tEd}{d\tD}
& \simeq \frac{\tG}{\pi} \biggl[ \frac{1}{\tD-\tEd}
  - \frac{2}{\tD+\tEd} + \frac{1}{\tD+\tU+\tEd} \, \biggr] , \\
\label{U:scaling}
\frac{d\tU}{d\tD}
&\simeq \frac{2\tG}{\pi} \biggl[ \frac{1}{\tD+\tEd}
   - \frac{1}{\tD-\tEd} \notag \\
&\qquad + \frac{1}{\tD-\tU-\tEd} - \frac{1}{\tD+\tU+\tEd} \, \biggr] .
\end{align}

Equations \eqref{Gamma:scaling}--\eqref{U:scaling} with initial
conditions $\tEd=\Ed$, $\tU=U$, and $\tG=\G$ represent the
main results of this section. The equations respect \ph\ symmetry
in that
\begin{equation}
\label{Ed+halfU:scaling}
\frac{d(\tEd+\half\tU)}{d\tD} \simeq \frac{2\tG}{\pi} \:
  \frac{\tEd+\half\tU}{(\tD-\half\tU)^2-(\tEd+\half\tU)^2} ,
\end{equation}
so bare couplings satisfying $\Ed=-\half U$ inevitably lead to rescaled
couplings that satisfy $\tEd=-\half\tU$. For $r=0$, Eqs.\
\eqref{Gamma:scaling}--\eqref{U:scaling} reproduce the scaling equations for
the metallic Anderson problem \cite{Hewson:93}, while for $r>0$ in the limit
$U\to\infty$ of extreme \ph\ asymmetry, Eqs.\ \eqref{Gamma:scaling}
and \eqref{Ed:scaling} reduce to ones presented previously
\cite{Gonzalez-Buxton:96} for pseudogapped systems.

Equation \eqref{Gamma:scaling} clearly has the solution
\begin{equation}
\label{tildeGamma}
\tG = (\tD/D)^r \; \G.
\end{equation}
Substituting this expression for $\tG$ into Eqs.\ \eqref{Ed:scaling} and
\eqref{U:scaling} creates a pair of coupled differential equations for $\tEd$
and $\tU$. Analytical or numerical integration of these differential equations
allows one to follow the evolution of the rescaled couplings under reduction
of $\tD$ until one of the following conditions is met, signaling entry into a
low-energy regime governed by a simpler effective model than the full pseudogap
Anderson model:

(1) If $\tEd,\,\tU+2\tEd>\tD>\tG$, the system lies in the empty-impurity
region of the \ASCm\ strong-coupling phase, in which the ground-state impurity
occupancy approaches zero. In this case, $T^*=\min(\tEd,\,\tU+2\tEd)$ sets
the scale for crossover into a low-energy regime of (for $r>0$, generalized)
Fermi-liquid behavior.

(2) If $-(\tU+\tEd),\,-(\tU+2\tEd)>\tD>\tG$, the system belongs in the
full-impurity region of the \ASCp\ strong-coupling phase, in which the
ground-state impurity occupancy approaches two. Here,
$T^*=\min(-(\tU+\tEd),\,-(\tU+2\tEd))$ marks crossover into the asymptotic
(generalized) Fermi-liquid regime.

(3) If $-\tEd,\,\tU+\tEd>\tD>\tG$, the system crosses over into an
intermediate-energy regime of local-moment behavior. (This \textit{regime}
is distinct from the LM \textit{phase}, which is defined by its ground-state
properties.) On entry to the LM regime, the empty and doubly occupied
impurity configurations are effectively frozen out, and one can perform a
generalization \cite{Gonzalez-Buxton:98} of the Schrieffer-Wolff transformation
\cite{Schrieffer:66} to map the pseudogap Anderson model to a pseudogap Kondo
model
\begin{equation}
\label{H_K}
H_K = H_{\band} + \frac{1}{N_k} \sum_{\bk,\bk'} \sum_{\s,\s'}
  c_{\bk\sigma}^{\dag} \biggl[ \frac{J}{2} \hat{\bm{S}} \cdot \bm{\s}_{\s\s'}
  + K \delta_{\s,\s'} \biggr] c_{\bk'\s'}^{\pdag} \,
\end{equation}
where $\H_{\band}$ is as given in Eq.\ \eqref{H_band} with the power-law
density of states specified in Eq.\ \eqref{rho:def}, $\hat{\bm{S}}$ is the
spin-$\half$ operator for the impurity, $\bm{\s}$ is a vector of Pauli matrices,
the (isotropic) exchange coupling $J$ satisfies
\begin{equation}
\label{J:SW}
\rho_0 J = \frac{2\tG}{\pi} \biggl(\frac{1}{|\tEd|}+\frac{1}{\tU+\tEd}\biggr),
\end{equation}
and the potential scattering $K$ satisfies
\begin{equation}
\label{K:SW}
\rho_0 K = \frac{\tG}{2\pi} \biggl(\frac{1}{|\tEd|}-\frac{1}{\tU+\tEd}\biggr).
\end{equation}
For metallic hosts ($r=0$), a system that reaches the LM regime always
lies in the strong-coupling phase of the Kondo model, which constitutes
another region of the strong-coupling phase of the Anderson model. In
pseudogap cases, by contrast, the asymptotic low-energy
behavior depends on the values of $J$ and $K$: the system may fall in one of
three Kondo phases that are associated with the SSC (for $K=0$), \ASCm\ (for
$K>0$), or \ASCp\ (for $K<0$) phases of the Anderson model; or it may fall
in the LM phase of both the Kondo and Anderson models, in which the
impurity retains a free two-fold spin degree of freedom down to absolute zero.
In any of these cases, the energy scale $T^*$ for crossover into the asymptotic
low-energy regime is generally much smaller than the scale
$\min(-\tEd,\,\tU+\tEd)$ for entry into the LM regime. On approach to a
strong-coupling ground state, $T^*$ coincides with the Kondo temperature
$T_K$.

(4) If $\tEd,\,-(\tU+\tEd)>\tD>\tG,\,\tU+2\tEd$ (a situation that
arises only if the bare $U$ is negative), then the system enters the
intermediate-energy local-charge regime. At this point, one can perform a
generalized Schrieffer-Wolff transformation to a pseudogap charge-Kondo model.
The system may lie in a strong-coupling phase of the charge-Kondo model (yet
another region of an Anderson-model strong-coupling phase) or in the
local-charge phase of both models, where the impurity retains a free two-fold
charge degree of freedom down to absolute zero. Similarly to the situation in
(3), the crossover to the asymptotic low-energy regime is characterized by
a scale $T^* \ll \min(\tEd,\,-(\tU+\tEd))$.

(5) If $\tG>\tD>|\tEd|$ and/or $\tG>\tD>|\tU+\tEd|$, then the system enters a
mixed-valence regime where the states $|\tilde{0}\rangle$,
$|\tilde{\sigma}\rangle$, and $|\tilde{2}\rangle$ are no longer all
well-defined. The scaling method is unable to determine whether the system
lies in the mixed-valence region of the strong-coupling phase,
or instead falls in the local-moment or local-charge phase \cite{Cheng-note}.

In the remainder of Sec.\ \ref{sec:Anderson}, we specialize to ranges of the
band exponent $r>0$ and the bare parameters $U$ (henceforth taken to be
positive, representing on-site Coulomb repulsion), $\Ed$, and $\G$ for which
it possible to make analytical predictions for the location of boundaries
between LM and strong-coupling phases. We compare these predictions with
results obtained using the non-perturbative numerical renormalization group
(NRG) method \cite{Krishna-murthy:80, Bulla:08}, as adapted to treat systems
containing a pseudogap density of states \cite{Bulla:97,Gonzalez-Buxton:98}.
Throughout the paper, we have set Wilson's discretization parameter
to $\Lambda=3$ and kept up to 600 many-body states after each iteration
of the NRG.

\subsection{Phase boundaries for $r>1$}
\label{subsec:r>1}

Analysis of band exponents in the range $r>1$ is simplified because Eq.\
\eqref{tildeGamma} means that $\tG/\tD = (\tD/D)^{r-1} (\G/D)$ decreases
monotonically under reduction of the half-bandwidth. In the physically most
relevant range $\G<D$, this decrease in $\tG/\tD$ rules out the possibility
of entry into the mixed-valence regime under condition (5) of Sec.\
\ref{subsec:poor man:A}.
Moreover, the decrease of $\tG$ is so rapid that any entry to the local-moment
regime and subsequent mapping to the pseudogap Kondo problem [via Eqs.\
\eqref{J:SW} and \eqref{K:SW}] will yield a sub-critical exchange coupling
that assigns the system to the local-moment phase \cite{Gonzalez-Buxton:98}.

Under these circumstances, the upper critical level energy
$\veps_{d,c}^+(\G,U)$ separating the \ASCm\ phase (in which $\tEd=\tD$ is
satisfied at sufficiently low $\tD$) from the LM phase (in which one eventually
reaches $\tEd=-\tD$) is effectively determined by the condition $\tEd(\tD=0)=0$
that places the fully renormalized impurity level precisely at the the Fermi
energy. This picture of the quantum phase transition as arising from a
renormalized level crossing is consistent with the observation of first-order
behavior for $r>1$ \cite{Ingersent:02,Fritz:04}. By \ph\ symmetry, the
boundary between the LM and \ASCp\ phases is at the lower critical
level energy $\veps_{d,c}^-=-U-\veps_{d,c}^+$ [see Fig.\
\ref{fig:PAM_phase}(b)].

The aforementioned boundary between the LM and \ASCm\ phases can be located by
performing an approximate integration of Eqs.\ \eqref{Ed:scaling} and
\eqref{U:scaling} using Eq.\ \eqref{tildeGamma}. For uniformity of
presentation, we express our result in the form of a critical hybridization
width $\G_c(U,\Ed)$. We will consider bare parameters satisfying
$0<-\Ed\ll U\!+\!\Ed, \,D$ and (for reasons that will become clear shortly)
$\G\ll(r-1)D$. Cases $U\ll D$ and $U\gg D$ will be considered separately.

\subsubsection{LM-\ASCm\ boundary for $0<-\Ed\ll U$ and $\G, \, U\ll D$}
\label{subsubsec:r>1:small-U}

If $U\ll D$, then so long as $|\tEd|,\,\tU+\tEd\ll\tD$, Eq.\
\eqref{U:scaling} can be approximated by
\begin{equation}
\label{U:scaling:small-U}
\frac{d\tU}{d\tD} \simeq \frac{4\tG\tU}{\pi\tD^2}
  = \frac{4\G}{\pi D^r} \, \tU \tD^{r-2},
\end{equation}
where use has been made of Eq.\ \eqref{tildeGamma}. This differential
equation can be integrated to yield
\begin{equation}
\label{tildeU:r>1:small-U}
\tU(\tD) \simeq U \exp\biggl[- \frac{4}{(r-1)\pi}
  \biggl( \frac{\G}{\ptD} - \frac{\tG}{\tD} \biggr) \biggr] ,
\end{equation}
which for $\G/D\ll (r-1)\pi$ describes a very weak downward renormalization
of $\tU$ with decreasing $\tD$.

During the same initial phase of the scaling, Eq.\ \eqref{Ed+halfU:scaling}
can be approximated by
\begin{equation}
\label{Ed:scaling:small-U}
\frac{d}{d\tD} (\tEd+\half\tU)
  \simeq \frac{2\G}{\pi D^r} \, (\tEd+\half\tU) \tD^{r-2},
\end{equation}
and hence
\begin{equation}
\label{tildeEd:r>1:small-U}
\tEd + \half \tU
\simeq (\Ed + \half U) \exp \biggl[ -\frac{2}{(r-1)\pi}
  \biggl( \frac{\G}{\ptD} - \frac{\tG}{\tD} \biggr) \biggr] .
\end{equation}
Equations \eqref{tildeU:r>1:small-U} and \eqref{tildeEd:r>1:small-U} imply that
\begin{equation}
\tEd - \Ed \simeq \frac{1}{4} \Bigl(1 - \frac{2\Ed}{U} \Bigr) \, (U - \tU) .
\end{equation}
In the case of present interest where $|\Ed|\ll U$, the level energy
scales upward in absolute terms by one-quarter the amount that the on-site
interaction scales down, but $\Ed$ experiences a much greater fractional
shift than $U$.

Equations \eqref{U:scaling:small-U}, \eqref{tildeU:r>1:small-U}, and
\eqref{tildeEd:r>1:small-U} remain valid until $(\tU+\tEd)/\tD$ rises to
approach unity, a condition that occurs [for the assumed ordering of the
bare parameters, and for the weak renormalization of $U$ that holds for
$\G\ll(r-1)D$] at $\tD=\tD_1\simeq U$, at which point
\begin{equation}
\label{tildeEd_1}
\tEdone \equiv \tEd(\tD_1)
  \simeq \Ed+\frac{\G U\bigl[ 1 - (U/D)^{r-1} \bigr]}{(r-1)\pi D} .
\end{equation}
In the regime $\tD<\tD_1$, the doubly occupied impurity configuration is
essentially frozen out. Now Eq.\ \eqref{Ed:scaling} can be approximated by
\begin{equation}
\label{Ed:scaling:large-U}
\frac{d\tEd}{d\tD} \simeq -\frac{\tG(\tD-3\tEd)}{\pi\tD^2}
 = -\frac{\G}{\pi D^r} \, (\tD-3\tEd) \tD^{r-2},
\end{equation}
which has the solution
\begin{equation}
\label{tildeEd:2a}
\tEd\simeq \tEdone
   + \frac{\G}{r\pi} \Biggl[ \biggl(\frac{\tD_1}{D}\biggr)^r
   - \biggl(\frac{\tD}{D}\biggr)^r \Biggr]
   \Biggr[1 + O\biggl( \frac{\tEdone}{U} \biggr)\Biggr].
\end{equation}
Using Eq.\ \eqref{tildeEd_1}, this gives
\begin{equation}
\label{tildeEd:2}
\tEd(\tD) \simeq \Ed + \frac{\G}{(r-1)\pi} \Biggl[\frac{U}{D}
  - \frac{1}{r} \biggl(\frac{U}{D}\biggr)^r
  - \frac{r-1}{r} \biggl(\frac{\tD}{D}\biggr)^r \Biggr].
\end{equation}

A more careful treatment of scaling over the range of $\tD$ in which
$|\tD-\tU-\tEd|\lesssim\tG$ [invalidating the nondegenerate perturbation
theory used to derive Eqs.\ \eqref{Gamma:scaling}--\eqref{U:scaling}] would
likely modify the numerical prefactor of $(U/D)^r \, \G$ on the right-hand
side of Eq.\ \eqref{tildeEd:2}. With this caveat, the equation should capture
the scaling of the impurity level energy until $|\tEd|/\tD$ grows to reach 1
at some reduced half-bandwidth $\tD_2$. For $\tD<\tD_2$, the system crosses over
into the low-energy regime of the \ASCm\ phase (for $\tEd>0$) or that of the
LM phase (for $\tEd<0$). The only exception occurs for a combination
of bare parameters that places the system precisely on the boundary between
the two phases, in which case $\tEd(\tD=0)=0$. Recalling that we are
considering cases $r>1$, Eq.\ \eqref{tildeEd:2} shows that the boundary
location $\veps_{d,c}^+(U,\G)$ is primarily determined by initial phase of
scaling ($\tD_1 < \tD < D$), and to leading order in $U/D$ and $\G/D$ satisfies
\begin{equation}
\label{Edc:r>1:1}
\veps_{d,c}^+ \simeq -\frac{\G U}{(r-1)\pi D} \, .
\end{equation}
This relation can be recast as
\begin{equation}
\label{Gammac:r>1:1}
\G_c \simeq (r-1)\pi D|\Ed|/U
\end{equation}
for $-U/2 \ll \Ed < 0$.

The phase boundary between the LM and \ASCm\ phases at $\G_c(U,\Ed)$ can be
determined to the desired accuracy by performing successive NRG runs to refine
the value of $\G_c$ using the method of bisection. At the end of each run,
the zero-temperature limit of $T\chi_{\imp}$ (temperature times the impurity
contribution to the static magnetic susceptibility) \cite{impurity-props,units}
is used to determine whether the system is in the LM phase
($T\chi_{\imp}\to 1/4$) or in the \ASCm\ phase ($T\chi_{\imp}\to 0$), and thus
to modify the range of $\G$ values within which $\G_c$ must lie.

\begin{figure}[t]
\centering
\includegraphics[width=\columnwidth]{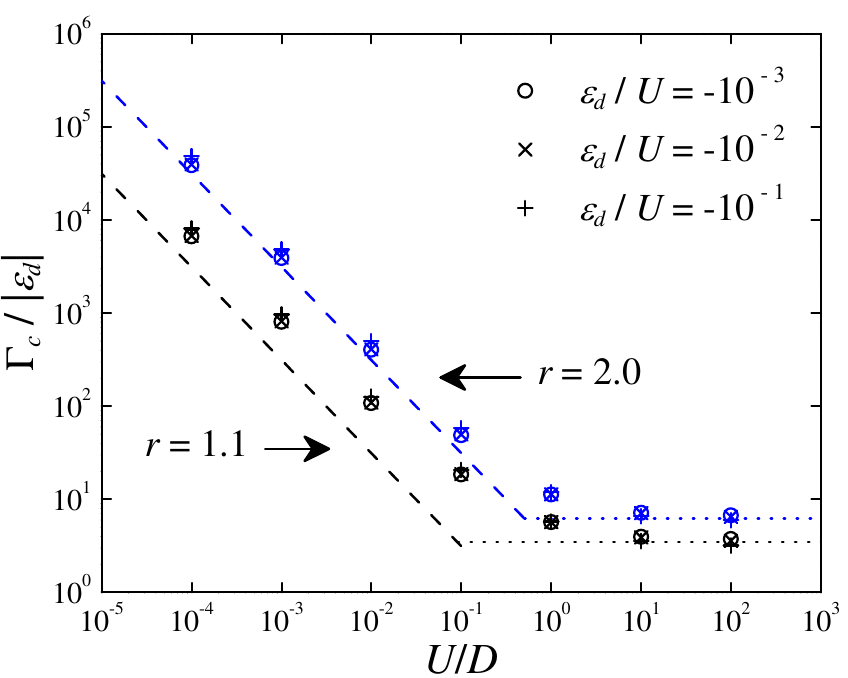}
\caption{\label{fig:r_gt_1}
Critical hybridization width plotted as $\Gamma_c/|\Ed|$ vs $U/D$, comparing
NRG data (symbols) with the scaling predictions of Eq.\ \eqref{Gammac:r>1:1}
(dashed lines) and Eq.\ \eqref{Gammac:U>>D} (dotted lines). Results are
for $r=1.1$ and $2$ and for three values of $|\Ed|/U$ shown in the
legend.}
\end{figure}

Figure \ref{fig:r_gt_1} shows the critical hybridization width plotted as
$\G_c(U, \Ed)/|\Ed|$ vs $U/D$ for $r = 1.1$ and $2$. NRG results (symbols)
calculated for three different fixed ratios $-\half\ll\Ed/U<0$ are compared
with the scaling expression in Eq.\ \eqref{Gammac:r>1:1} (dashed lines).
The numerics confirm the predicted linear dependence of $\G_c$ on $\Ed$. The
$U^{-1}$ variation of $\G_c$ is also well supported for $r=2$, but for $r=1.1$
is only approached in the limit of very small $U/D$. This deviation from the
scaling theory likely arises because the latter band exponent lies close to
the value $r=1$ that acts as an upper critical dimension for the pseudogap
Anderson and Kondo models and at which there are logarithmic corrections to
simple power-law behaviors \cite{Cassanello:96,Ingersent:02,Fritz:04}.
Results for $1.1<r<2$ (not shown) indicate that increasing $r$ leads to a
continuous improvement in the accuracy with which Eq.\ \eqref{Gammac:r>1:1}
reproduces the NRG data.

\subsubsection{LM-\ASCm\ boundary for $0<-\Ed, \G\ll D\ll U$}
\label{subsubsec:r>1:large-U}

If the bare parameters of the Anderson Hamiltonian instead satisfy $U\gg D$,
then Eq.\ \eqref{Ed:scaling:large-U} applies from the outset of scaling, and
$\tEd$ satisfies Eq.\ \eqref{tildeEd:2a} with $\tD_1=D$ and $\tEdone=\Ed$,
i.e.,
\begin{equation}
\label{tildeEd:large-U}
\tEd(\tD) \simeq \Ed + \frac{\G}{r\pi}
  \Biggl[ 1 - \biggl(\frac{\tD}{D}\biggr)^r \Biggr] .
\end{equation}
Now the condition $\tEd(\tD=0)=0$ places the LM-\ASCm\ phase boundary at
\begin{equation}
\label{Gammac:U>>D}
\G_c \simeq r\pi|\Ed| .
\end{equation}

Figure \ref{fig:r_gt_1} compares the prediction of Eq.\ \eqref{Gammac:U>>D}
(dotted lines) with NRG results obtained for $U \ge D$. The scaling approach
reproduces the numerical results very well for $U\gtrsim 10D$, and (in
contrast to the behavior found for $U\ll D$) there is no significant
difference between $r=1.1$ and $r=2$ in the accuracy of the analytical
results.

\subsection{Phase boundaries for $0<r<1$}
\label{subsec:r<1}

For $r<1$, Eq.\ \eqref{tildeGamma} implies that
$\tG/\tD = ( D / \tD )^{1-r} (\G/D) \ge \G/D$.
The system flows to mixed valence [under condition (5) in Sec.\
\ref{subsec:poor man:A}] at a reduced half-bandwidth
\begin{equation}
\tD_{\G} = \tG(\tD_{\G}) = (\G/D)^{1/(1-r)} \, D
\end{equation}
provided that $|\tEd(\tD_{\G})|$ and $|\tU(\tD_{\G})+\tEd(\tD_{\G})|$
both remain smaller than $\tD_{\G}$. However, the system flows to a
different low-energy regime if $|\tEd|/\tD$ or $|\tU+\tEd|/\tD$ reaches $1$
at some $\tD > \tD_{\G}$.

\subsubsection{LM-SSC boundary for $\G, \, U\ll D$}
\label{subsubsec:r<1:symm}

We first consider cases $\Ed=-\half U$ where the system exhibits strict
\ph\ symmetry, and focus on the universal (large-bandwidth)
limit $\G, U \ll D$.

So long as $\half\tU\ll\tD$, Eq.\ \eqref{U:scaling} can again be approximated
by Eq.\ \eqref{U:scaling:small-U}, which can be integrated to yield
\begin{equation}
\label{tildeU:r<1:small-U}
\tU(\tD) \simeq U \exp\biggl[- \frac{4}{(1-r)\pi}
  \biggl( \frac{\tG}{\tD} - \frac{\G}{\ptD} \biggr) \biggr] .
\end{equation}
Equation \eqref{tildeU:r<1:small-U} can be reexpressed as
\begin{equation}
\label{tUbytD:r<1:small-U}
(\tU/2\tD)^{1-r} \simeq \tx \, e^{-\gamma \, \tx}
\end{equation}
in terms of new variables
\begin{equation}
\label{tildex:def}
\tx(\tD)
  = \biggl(\frac{U}{2\tD}\biggr)^{1-r} \exp \biggl(\frac{4\G}{\pi D}\biggr)
  \ge x \equiv \tx(D)
\end{equation}
and
\begin{equation}
\label{gamma:def}
\gamma = \biggl(\frac{2D}{U}\biggr)^{1-r} \: \biggl(\frac{4\G}{\pi D}\biggr)
         \exp\biggl(-\frac{4\G}{\pi D}\biggr)
\end{equation}
that allow Eq.\ \eqref{tildeGamma} to be recast exactly in the form
\begin{equation}
\label{tGbytD}
\tG/\tD = \frac{\pi}{4} \, \gamma \, \tx .
\end{equation}

Equation \eqref{tUbytD:r<1:small-U} shows that with increasing $\tx$ (or
decreasing $\tD$), $\tU/2\tD$ initially rises, before peaking at $\tx=1/\gamma$,
and then dropping off exponentially for $\tx\gg 1/\gamma$. The system will enter
its local-moment regime [under condition (3) in Sec.\ \ref{subsec:poor man:A}]
if there exists a reduced half-bandwidth $\tD_U>\tD_{\G}$ such that
$\tU(\tD_U)/2\tD_U=1$. The approximate scaling equation
\eqref{U:scaling:small-U} is valid only so long as $\tU/\tD\lesssim 1$.
Equation \eqref{U:scaling} predicts that $\tU$ experiences a stronger downward
renormalization once $\tU/\tD$ approaches $2$, a range in which the
nondegenerate perturbation theory used to derive Eqs.\
\eqref{Gamma:scaling}--\eqref{U:scaling} also begins to break down. However, in
this range of $\tU/2\tD$, physically one expects renormalization to slow to a
halt as charge fluctuations are progressively frozen out. Therefore, in the
spirit of Haldane \cite{Haldane:78}, we apply Eq.\ \eqref{tUbytD:r<1:small-U}
all the way to the point where $\tU(\tD)/2\tD=1$, and we seek $\tx_U$
defined to be the smallest solution of
\begin{equation}
\label{LM-cond:symm}
\tx \, e^{-\gamma \, \tx} = 1.
\end{equation}

\begin{figure}[t]
\centering
\includegraphics[width=\columnwidth]{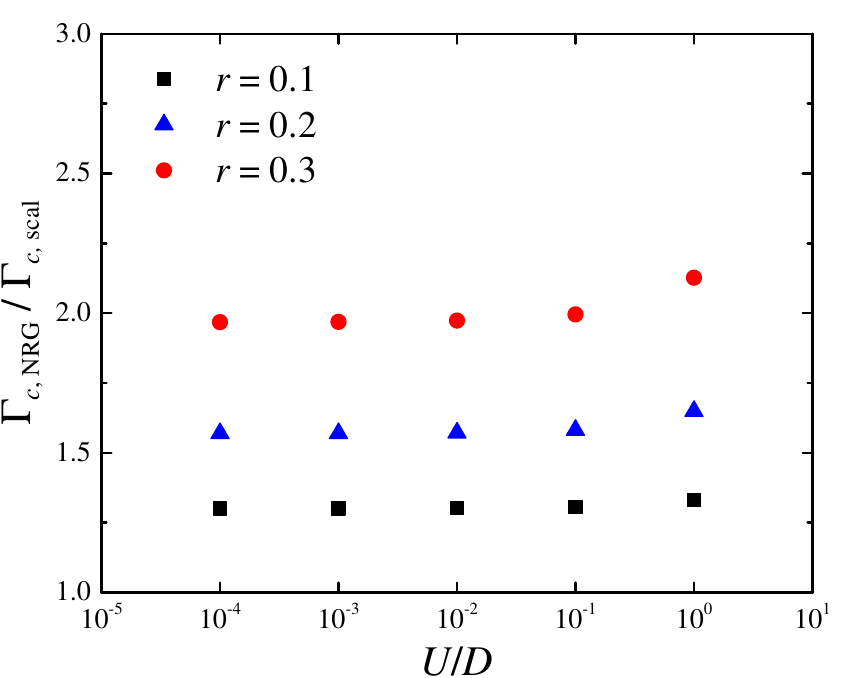}
\caption{\label{fig:r<1:PHSym}
Ratio $\Gamma_{c,\text{NRG}}\,/\,\G_{c,\text{scal}}$ of the critical
hybridization width found using NRG to the scaling prediction given by
Eq.\ \eqref{Gamma_c:r<1}, plotted as a function of $U/D$ for $\Ed=-U/2$ and
band exponents $r=0.1$, $0.2$, and $0.3$.}
\end{figure}

For $\gamma > 1/e$, Eq.\ \eqref{LM-cond:symm} has no real solution, so the
system necessarily crosses over to mixed valence for $\tD\lesssim\tD_{\G}$.
For $0\le \gamma\le 1/e$, by contrast, Eq.\ \eqref{LM-cond:symm} has a
solution $\tx_U(\gamma)$ satisfying $1\le\tx_U \le e \le \gamma^{-1}$.
Since $\gamma \tx_U<1$, Eq.\ \eqref{tGbytD} gives $\tG(\tD_U)<\tD_U$, meaning
that at $\tD=\tD_U$ the system satisfies condition (3) for crossover into its
local-moment regime. Equation \eqref{tildex:def} gives
\begin{equation}
\frac{\tU}{U}
  = \frac{2\tD_U}{U}
  = \biggl[\frac{\exp(4\G/\pi D)}{\tx_U(\gamma)}\biggr]^{1/(1-r)}
  \ge e^{-1/(1-r)}
\end{equation}
since $\tx_U \le e$. This implies, at least for $r\ll\half$, that the rescaled
on-site interaction $\tU(\tD_U)$ remains of the same order as $U$.

A Schrieffer-Wolff transformation performed at $\tD=\tD_U$ yields a pseudogap
Kondo model with [Eqs.\ \eqref{J:SW} and \eqref{K:SW}]
\begin{equation}
\label{SW:r<1}
\rho_0 J = \frac{8\tG(\tD_U)}{\pi\tU(\tD_U)} = \gamma \, \tx_U,
  \qquad \rho_0 K = 0.
\end{equation}
It is known that for $\rho_0 K = 0$, the critical exchange coupling $J_c$
separating the Kondo ($J>J_c$) and LM ($J<J_c$) phases satisfies
$\rho_0 J_c = f(r)$ where $f(r) \simeq r(1+r/2)$ for $r\ll \half$ (Refs.\
\onlinecite{Withoff:90} and \onlinecite{Ingersent:96}) and $f(r)\to\infty$ for
$r\to \half^-$ (Ref.\ \onlinecite{Ingersent:96a}). Combining this information
with Eq.\ \eqref{SW:r<1}, one arrives at the prediction that the boundary
between the LM and SSC phases is determined by the condition
$\gamma_c \, \tx_U(\gamma_c) = f(r)$. Then, Eq.\ \eqref{LM-cond:symm} gives
$\tx_c \equiv \tx_U(\gamma_c)=\exp[f(r)]$ and, hence,
$\gamma_c = f(r) / \tx_c = f(r) \, \exp[-f(r)]$. This means that the LM phase
occupies the parameter range $U>U_c(\G)$, where
\begin{equation}
\label{U_c:r<1}
U_c = 2 D \biggl\{ \frac{\exp[f(r)]}{f(r)} \: \frac{4\G}{\pi D} \:
      \exp\biggl(-\frac{4\G}{\pi D}\biggr) \biggr\}^{1/(1-r)} .
\end{equation}

For $\G\ll D$, one can invert Eq.\ \eqref{U_c:r<1} to deduce that the LM
phase occupies the parameter range $\G<\G_c(U)$, with
\begin{equation}
\label{Gamma_c:r<1}
\G_c \simeq D \, \frac{\pi f(r)}{4 \exp[f(r)]} \:
   \biggl(\frac{U}{2D}\biggr)^{1-r}.
\end{equation}
A $U^{1-r}$ variation of $\G_c$ was found previously using the local-moment
approach \cite{Logan:00}, which yields a closed-form expression for
$r\to 0^+$ that is identical to the corresponding limit of Eq.\
\eqref{Gamma_c:r<1}.

Figure \ref{fig:r<1:PHSym} plots the ratio of the critical hybridization
width $\G_{c,\mathrm{NRG}}$ found using the NRG to the scaling prediction
$\G_{c,\mathrm{scal}}$ given by Eq.\ \eqref{Gamma_c:r<1}. For band exponents
$r=0.1$, $0.2$, and $0.3$, this ratio is well converged for
$U/D \lesssim 0.1$, implying that the scaling analysis correctly captures
the $U^{1-r}$ dependence of $\G_c$ at the LM-SSC phase boundary. The absolute
value of $\G_{c,\mathrm{NRG}}/\G_{c,\mathrm{scal}}$ falls as $r$ decreases,
and seems likely to approach unity as $r\rightarrow 0^+$. We infer that Eq.\
\eqref{Gamma_c:r<1} describes the NRG results apart from a multiplicative
correction factor that depends solely on the band exponent $r$.

\subsubsection{Kondo-mixed valence crossover for $\G, \, U\ll D$}
\label{subsubsec:r<1:symm:MV}

\begin{figure}[t]
\centering
\includegraphics[width=\columnwidth]{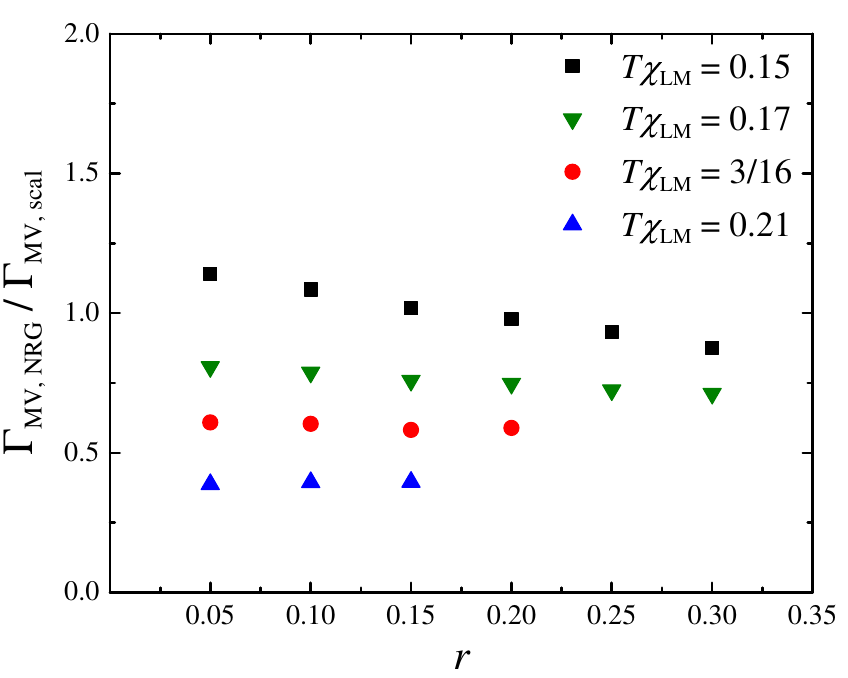}
\caption{\label{fig:r<1:PH-MV}
Ratio $\Gamma_{\text{MV,NRG}}\,/\,\G_{\text{MV,scal}}$ of the mixed-valence
threshold hybridization width found using NRG and that given by poor man's
scaling. Results for fixed  $U=-2\Ed=10^{-4}D$ are plotted vs band exponent $r$
with $\Gamma_{\text{MV,NRG}}$ defined using four different values of
$T\chi_{\LM}$ (see text for details). For $T\chi_{\LM}=3/16$ [$0.21$], it proves
impossible to find Kondo-region behavior for $r\ge 0.2$ [$r\ge 0.15$].}
\end{figure}

Poor man's scaling not only can find the LM-SSC phase boundary at $\G=\G_c$,
but also can locate a crossover within the SSC phase at $\G=\G_{\MV}$ between
a Kondo region, in which only the singly occupied impurity configurations
have significant occupation at low temperatures, and a mixed-valence region
also having significant ground-state occupancy of the empty and/or doubly
occupied impurity configuration(s). We have seen [after Eq.\
\eqref{LM-cond:symm}] that the system reaches mixed valence for $\gamma>e^{-1}$
(equivalent to $\G>\G_{\MV}$, a mixed-valence threshold hybridization),
and argued [before Eq.\ \eqref{U_c:r<1}] that it enters the LM phase for
$\gamma<\gamma_c = f(r) \, \exp[-f(r)]$. Therefore, the system exhibits fully
developed Kondo physics (i.e., enters its local-moment regime at intermediate
values of $\tD$ and then crosses over to the SSC ground state for
$\tD\lesssim T_K$) only for $\gamma_c<\gamma \le e^{-1}$, equivalent to the
condition
$\G_c < \G \le \G_{\MV}$ with
\begin{equation}
\label{Gamma_MV}
\G_{\MV} \simeq \G_c \exp[f(r)-1] / f(r) .
\end{equation}
The Kondo region has a width $\Delta\G=\G_{\MV}-\G_c$ that narrows rapidly
with increasing $r$ and turns out to be restricted to $\Delta\G \lesssim \G_c/4$
for $r \gtrsim \frac{1}{3}$. For band exponents in the range $\frac{1}{3}
\lesssim r < \half$, the SSC phase can be accessed only from mixed valence,
while for $r\ge\half$ this phase disappears altogether
\cite{Bulla:97,Gonzalez-Buxton:98}.

Within the NRG approach, we can define the mixed-valence threshold hybridization
width by examining the temperature dependence of the impurity contribution
to the magnetic susceptibility $\chi_{\imp}$. We can identify the Anderson model
as being in its local-moment regime if $T\chi_{\imp}>T\chi_{\LM}$ where
$T\chi_{\LM}$ is a (somewhat arbitrary) cutoff chosen to lie between the value
$T\chi_{\imp}=1/4$ corresponding to a free spin-$\half$ degree of freedom
and the high-temperature or mixed-valent limiting value $T\chi_{\imp}=1/8$.
With this criterion, the system is in the Kondo region of the SSC phase if
with decreasing $T$, $T\chi_{\imp}$ first rises above $T\chi_{\LM}$ before
dropping towards its SSC value \cite{Gonzalez-Buxton:98} of $r/8$. We therefore
define $\G_{\text{MV,NRG}}$ as the smallest hybridization width $\Gamma$ for
which $T\chi_{\imp}<T\chi_{\LM}$ at all temperatures.

Figure \ref{fig:r<1:PH-MV} shows the ratio
$\G_{\text{MV,NRG}}\,/\,\G_{\text{MV,scal}}$ between the mixed-valence
threshold coupling found using NRG and the scaling prediction of Eq.\
\eqref{Gamma_MV}. The ratio is plotted vs band exponent $r$ for fixed
$U/D=10^{-4}$ and four different cutoffs: $T\chi_{\LM} = 0.15, 0.17, 3/16$,
and $0.21$. As one would expect, increasing the value of $T\chi_{\LM}$ creates
a more stringent criterion for the identification of Kondo physics, reduces the
range of exponents $r$ over which Kondo-region behavior is found, and for
given $r$ reduces the value of $\G_{\MV}$. However, the ratio
$\G_{\text{MV,NRG}}/\G_{\text{MV,scal}}$ is nearly independent of $r$ except
in the case $T\chi_{\LM} = 0.15$.
This confirms that the condition for reaching mixed valence is correctly
captured by Eq.\ \eqref{Gamma_MV} apart from a multiplicative factor that
depends on the value of the cutoff $T\chi_{\LM}$.

\subsubsection{LM-\ASCm\ boundary for $0<-\Ed\ll U+\Ed$ and $\G\ll D$}
\label{subsubsec:r<1:asymm}

We now turn to the limit $0<-\Ed\ll U\!+\!\Ed, \,D$ of strong \ph\
asymmetry on the impurity site. In order to locate the boundary between the
LM and \ASCm\ phases, we will perform an approximate integration of Eqs.\
\eqref{Ed:scaling} and \eqref{U:scaling} using Eq.\ \eqref{tildeGamma}.
We first consider $\G, \, U\ll D$. The situation where $\G\ll D\ll U$ will be
considered at the end of the section.

So long as $|\tEd|,\,\tU+\tEd\ll\tD$, Eq.\ \eqref{U:scaling} can once more be
approximated by Eq.\ \eqref{U:scaling:small-U}, leading to Eq.\
\eqref{tildeU:r<1:small-U}, and Eq.\ \eqref{Ed+halfU:scaling} can again be
approximated by Eq.\ \eqref{U:scaling:small-U}, which yields
\begin{equation}
\label{tildeEd:r<1:small-U}
\tEd + \half \tU
\simeq (\Ed + \half U) \exp \biggl[ -\frac{2}{(1-r)\pi}
  \biggl( \frac{\tG}{\tD} - \frac{\G}{\ptD} \biggr) \biggr] .
\end{equation}
Equations \eqref{tildeU:r<1:small-U} and \eqref{tildeEd:r<1:small-U} imply that
\begin{equation}
\tEd + \half \tU
\simeq (\Ed + \half U) \sqrt{\tU/U} .
\end{equation}

Equations \eqref{U:scaling:small-U}, \eqref{tildeU:r<1:small-U}, and
\eqref{tildeEd:r<1:small-U} remain valid until either $\tG/\tD$ reaches $1$
at $\tD=\tD_{\G}$ or $(\tU+\tEd)/\tD$ reaches $1$ at $\tD=\tD_1$.
By writing $\tU+\tEd = \half\tU + (\tEd+\half\tU)$, then employing
Eqs.\ \eqref{tildeU:r<1:small-U}, \eqref{tUbytD:r<1:small-U}, and
\eqref{tildeEd:r<1:small-U}, the latter condition can be recast as
\begin{equation}
\label{LM-cond:asymm}
\tx e^{-\gamma\tx/2} \bigl[ \eta
  + e^{-\gamma\tx/2(1\!-\!r)} \bigr]^{1-r} = 1,
\end{equation}
with $\tx$ and $\gamma$ as defined in Eqs.\ \eqref{tildex:def} and
\eqref{gamma:def}, and
\begin{equation}
\eta = \biggl( 1 + \frac{2\Ed}{U} \biggr)
       \exp \biggl[-\frac{2\G}{\pi(1-r)D}\biggr] \simeq 1 .
\end{equation}
Given Eq.\ \eqref{tGbytD},
the conditions $\tG(\tD_1)<\tU(\tD_1)+\tEd(\tD_1)=\tD_1$ are satisfied
provided that Eq.\ \eqref{LM-cond:asymm} has a real solution
$\tx=\tx_1(\gamma)<4/\pi\gamma$. Such solutions exist for
$\gamma\lesssim\gamma_{\max}(r)
  = (4/\pi) \, e^{-2/\pi} \bigl[ \eta + e^{-2/\pi(1-r)} \bigr]^{1-r}$.
For $\eta=1$, there is a monotonic decrease in $\gamma_{\max}$
with increasing $r$, from $\gamma_{\max}(0^+) \simeq 1.030$ to
$\gamma_{\max}(1^-) \simeq 0.674$, while the solution to Eq.\
\eqref{LM-cond:asymm} satisfies
$2^{-(1-r)}\le \tx_1 \le 4/\pi\gamma_{\max}<1.89$.

In the regime $\tD<\tD_1$, entered with $\tEd=\tEdone$, the
doubly occupied impurity configuration is essentially frozen out. Now
Eq.\ \eqref{Ed:scaling} can be approximated by Eq.\
\eqref{Ed:scaling:large-U}, again yielding Eq.\ \eqref{tildeEd:2a}.
This second phase of the scaling continues until one of the following
conditions is met:
(a) $\tEd=\tD$, signaling crossover into the empty-impurity region of the
\ASCm\ phase;
(b) $\tG=\tD$, marking entry into the mixed-valence region of the \ASCm\
phase;
(c) $\tEd=-\tD$, marking entry into the local-moment regime.
In case (c), the system may be mapped onto the pseudogap Kondo Hamiltonian
described by Eqs.\ \eqref{H_K}--\eqref{K:SW}, which may lie in (c)(i) the
\ASCm\ phase, or (c)(ii) the LM phase.
Integrating the poor man's scaling equations with sufficient accuracy
to distinguish among all these possibilities is in general a formidable
challenge.

Progress on locating the LM-\ASCm\ phase boundary can be made in the limit
$\gamma\ll 1$ of very weak impurity-band hybridization, where
$\tD_1\simeq U+\Ed$ and $\tx_1=(1+\eta)^{-(1-r)}[1+O(\gamma)]$. Focusing for
simplicity on $\eta\to 1$, one finds
\begin{equation}
\tEdone \simeq \Ed + \frac{\G}{(1-r)\pi} \biggl( \frac{U}{D} \biggr)^r ,
\end{equation}
and hence [via Eq.\ \eqref{tildeEd:2a}]
\begin{equation}
\tEd = \Ed + \frac{\G}{r\pi} \biggl[ \frac{1}{1-r}
\biggl(\frac{U}{D}\biggr)^r - \biggl(\frac{\tD}{D}\biggr)^r \biggr] .
\end{equation}
In this limit of small $\gamma$, one expects only a small fractional change in
the bare level energy $\Ed$ to be required to drive the system from
case (a) to case (c)(ii) of the previous paragraph. Under these circumstances,
just as was done with greater rigor for $r>1$, one can approximate the location
of the phase boundary by the condition $\tEd(\tD=0)=0$, leading to
\begin{equation}
\label{Edc:r<1:1}
\veps_{d,c}^+ \simeq -\frac{\G}{r(1-r)\pi} \, \biggl( \frac{U}{D} \biggr)^r \, .
\end{equation}
Equation \eqref{Edc:r<1:1} can be inverted such that the system is in the LM
phase if $\G<\G_c$, where the critical coupling is given by
\begin{equation}
\label{Gammac:r<1}
\G_c=r(1-r)\pi \, |\veps_{d}|  \bigg( \frac{U}{D} \bigg)^{-r}.
\end{equation}

For $U\gg D$, the evolution of $\tEd$ with $\tD$ is as described by Eq.\
\eqref{tildeEd:large-U}. For $\gamma\ll 1$, arguments similar to those given
at the end of the previous section lead to the conclusion that the LM-\ASCm\
boundary is given by Eq.\ \eqref{Gammac:U>>D}.

\begin{figure}[t]
\centering
\includegraphics[width=\columnwidth]{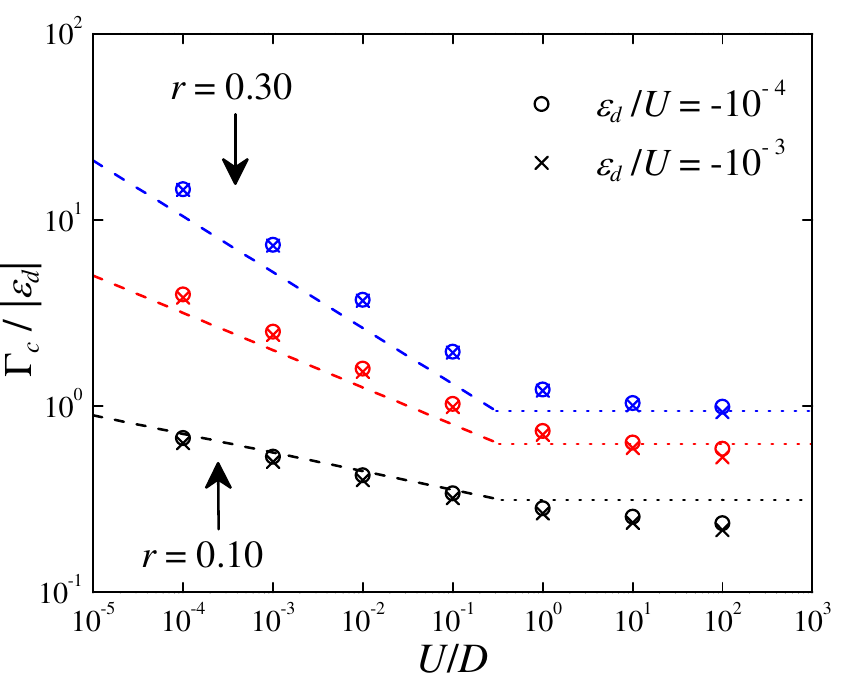}
\caption{\label{fig:r<1:PHAsym}
Critical coupling $\G_c$ found using NRG (symbols) plotted as $\G_c/|\Ed|$
against $U/D$ for $\veps_d/U=-10^{-4}$ and $-10^{-3}$ and for $r=0.1$, $0.2$
and $0.3$. Also plotted are the poor man's scaling prediction of
Eq.\ \eqref{Gammac:r<1} for $U\ll D$ (dashed lines) and
Eq.\ \eqref{Gammac:U>>D} for $U\gg D$ (dotted lines).}
\end{figure}

Figure \ref{fig:r<1:PHAsym} shows the critical hybridization width plotted as
$\G_{c}/|\veps_d|$ vs $U/D$ for band exponents $r=0.1$, $0.2$ and $0.3$ and for
two values of the ratio $|\veps_d|/U$ listed in legend. NRG data (symbols) are
compared with the poor man's scaling predictions. For $U\ll D$, $\G_c/|\veps_d|$
exhibits a $(U/D)^{-r}$ dependence that is described very well by Eq.\
\eqref{Gammac:r<1} (dashed lines) apart from an overall multiplicative factor
that grows with increasing $r$. For $U\gg D$, $\G_c/|\veps_d|$ is almost
(but not quite) a constant as predicted by Eq.\ \eqref{Gammac:U>>D} (dotted
lines). These behaviors show that the poor man's scaling approach provides a
good account of the phase boundary in the limit of strong \ph\ asymmetry
on the impurity site.

\section{Anisotropic Power-Law Kondo Model}
\label{sec:Kondo}

In this section, we present a poor man's scaling analysis of the phase
boundary between the Kondo and local-moment (LM) phases of the Kondo model
with distinct longitudinal and transverse spin-flip couplings between the
impurity and a power-law-vanishing or power-law-diverging density of
states. The model is described by the Hamiltonian
\begin{equation}
\label{H_K:aniso}
\H_K = \H_{\band} + J_z \hat{S}_z \hat{s}_z + \frac{1}{2} J_{\!\p}
  \bigl( \hat{S}^+ \hat{s}^- + \hat{S}^- \hat{s}^+ \bigr) ,
\end{equation}
where $\H_{\band}$ is as given in Eq.\ \eqref{H_band} with the
density of states specified in Eq.\ \eqref{rho:def}, and $\hat{\bm{S}}$ and
$\hat{\bm{s}}=N_k^{-1} \sum_{\bk,\bk'} \sum_{\s,\s'} c_{\bk\sigma}^{\dag}
\half \bm{\s}_{\s\s'} c_{\bk'\s'}^{\pdag}$ (with $\bm{\s}$ being a vector
of Pauli matrices) are, respectively, the spin-$\half$ operators for the
impurity and for conduction band electrons at the impurity site. The properties
of the model are invariant under $J_{\!\p}\to -J_{\!\p}$, but for notational
simplicity we will consider only $J_{\!\p}\ge 0$.

Our focus is primarily on pseudogap cases $r>0$, which can arise, for example,
due to the low-temperature freeze-out of charge fluctuations in the
Anderson-Holstein model with a power-law density of states \cite{Cheng:13}.
However, in Sec.\ \ref{subsec:divergent:K} we briefly consider the range
$-1<r<0$ describing bands with a generalized Van Hove singularity at the Fermi
energy \cite{Vojta:02,Mitchell:13}.

\subsection{Poor man's scaling equations}
\label{subsec:scaling:K}

By generalizing Anderson's poor man's scaling treatment of the conventional
($r=0$) Kondo problem \cite{Anderson:70b}, it is straightforward to extend
Withoff and Fradkin's analysis of the pseudogap Kondo problem to anisotropic
exchange. Under progressive reduction of the half-bandwidth from $D$ to
$\tD=D e^{-l}$, the exchange couplings $(J_z, \, J_{\!\p})$ evolve to
$(\tJ_z, \, \tJ_{\p})$ according to
\begin{subequations}
\label{aniso:scaling}
\begin{equation}
\label{J_z:scaling}
\frac{d\tJ_z}{dl}
  = -r \tJ_z + \rho_0 \tJ_{\p}^2,
\end{equation}
and
\begin{equation}
\label{J_p:scaling}
\frac{d\tJ_{\p}}{dl}
  = -r \tJ_{\p} + \rho_0 \tJ_z \tJ_{\p}.
\end{equation}
\end{subequations}
On the right-hand side of each of these equations, the first term reflects the
change in the density of states at the band edge (a single-particle effect),
while the second term reflects the lowest-order many-body effects and is
independent of $r$. These equations neglect all contributions beyond
second-order in the exchange, and are therefore restricted in validity to
situations where $|\rho_0 \tJ_z| \ll 1$ and $\rho_0 \tJ_{\!\p} \ll 1$.

Equations \eqref{aniso:scaling} can be combined to obtain
\begin{equation}
\frac{d}{dl} \bigl( \tJ_z^2 - \tJ_{\p}^2 \bigr)
  = -2r \bigl( \tJ_z^2 - \tJ_{\p}^2 \bigr) ,
\end{equation}
which can be integrated to yield
\begin{equation}
\label{aniso-irrel}
\tJ_z^2 - \tJ_{\p}^2 = \bigl(J_z^2 - J_{\!\p}^2\bigr) e^{-2rl},
\end{equation}
One sees that exchange anisotropy is irrelevant for $r>0$ (pseudogapped
systems), marginal for $r=0$ (conventional metals), and relevant for $r<0$
(describing a power-law divergence of the host density of states at the Fermi
energy).
Equation \eqref{aniso-irrel} can be inserted into Eq.\ \eqref{J_z:scaling}
to obtain
\begin{equation}
\label{J_z:ode-scaling}
\frac{d\tJ_z}{dl} = -r \tJ_z + \rho_0\tJ_z^2
  - \rho_0\bigl(J_z^2 - J_{\!\p}^2\bigr) e^{-2rl}.
\end{equation}

After the completion of the work reported in this paper, we learned of a
recent poor man's scaling formulation of the power-law Kondo model with
a more general anisotropic exchange coupling $J_x \hat{S}_x \hat{s}_x +
J_y \hat{S}_y \hat{s}_y + J_z \hat{S}_z \hat{s}_z$ \cite{Kogan:17}. For
the case $J_x=J_y=J_{\perp}$ considered here, the scaling equations of
Ref.\ \onlinecite{Kogan:17} reduce to Eqs.\ \eqref{aniso:scaling} and
yield scaling trajectories fully equivalent in appearance to those
plotted in Figs.\ \ref{fig:flows:r>0} and \ref{fig:flows:r=-0.1} of this
paper.

\subsection{Pseudogapped density of states}
\label{subsec:pseudogap:K}

For $r>0$, Eqs.\ \eqref{aniso:scaling} have two stable fixed points, both
isotropic as expected from Eq.\ \eqref{aniso-irrel}: the weak-coupling or
LM fixed point $\tJ_z=\tJ_{\p}=0$, and the strong-coupling or
Kondo fixed point $\tJ_z=\tJ_{\p}=\infty$
(which lies beyond the range of validity of the equations but is known to
exist from nonperturbative studies). There is also a critical fixed point
$\rho_0 \tJ_z = \rho_0 \tJ_{\p} = r$ that lies on the boundary
between the basins of attraction of the stable fixed points.
The goal of this subsection is to map out the location of this boundary away
from the point of SU(2) spin symmetry. In light of Eq.\ \eqref{aniso-irrel},
it is clear that any starting point on the boundary flows under Eqs.\
\eqref{aniso:scaling} to the isotropic critical point first identified by
Withoff and Fradkin \cite{Withoff:90}, which therefore governs the
low-energy physics.

For $J_z\ne 0$, one can factorize out the variation of $\tJ_z$
arising from pure density-of-states effects [i.e., the effect of the
$-r\tJ_z$ term on the right-hand-side of Eq.\ \eqref{J_z:ode-scaling}]
through the substitution
\begin{equation}
\label{tj:def}
\tJ_z = \tj(l) \, J_z \, e^{-rl} ,
\end{equation}
which converts Eq.\ \eqref{J_z:ode-scaling} to
\begin{equation}
\label{tj:scaling}
\frac{d\tj}{dl}
  = \Bigl[ \tj^{\,2} - 1 + \bigl(J_{\!\p}/J_z\bigr)^2 \Bigr] \rho_0 J_z \, e^{-rl}
\end{equation}
with the initial condition $\tj(0)=1$.
For any antiferromagnetic bare exchange $J_z>0$ and any $J_{\!\p}>0$, Eq.\
\eqref{tj:scaling} yields $d\tj/dl \ge 0$ so that $\tj>1$ for all $l>0$.
If $\tj$ remains finite as $l\to\infty$, then $\tJ_z$ vanishes as $\tD\to 0$
and the system must lie in the LM phase. On the other hand, we can associate
the divergence of $\tj$ at some value $l=l_K$ with entry into the Kondo regime
around temperature $T_K=De^{-l_K}$. The boundary between the two phases is
determined by the divergence of $\tj(l)$ only at $l=\infty$. For a
ferromagnetic bare exchange $J_z<0$, any $J_{\!\p}\ne 0$ is sufficient to
ensure that $\tj<1$ for all $l>0$. In this case, the system enters the Kondo
regime if $\tj$ changes sign and reaches $-\infty$ for some finite $l_K$.

For the purposes of more detailed analysis, it proves convenient to parametrize
the anisotropy of the bare exchange couplings in terms of the variable
\begin{equation}
\label{alpha:def}
\alpha = \sqrt{\bigl|(J_{\!\p}/J_z)^2 -1\bigr|}
  \;\; \text{sgn} \bigl[(J_{\!\p}/J_z)^2 -1\bigr],
\end{equation}
which can range from $-1$ (for $J_{\!\p}=0$) to $0$ (for $J_{\!\p}=|J_z|$) to
$+\infty$ (for $J_{\!\p}\gg|J_z|$).
Then Eq.\ \eqref{tj:scaling} can be rewritten
\begin{equation}
\label{tj:scaling1}
\frac{d\tj}{dl} = \bigl(\tj^2 + \alpha |\alpha|) \, \rho_0 J_z \, e^{-rl}.
\end{equation}
Solutions of this equation will be examined in the next two subsections.

\subsubsection{Easy-plane anisotropy}

In cases where $J_{\!\p}>|J_z|>0$, $\alpha$ defined in Eq.\ \eqref{alpha:def} is
positive and Eq.\ \eqref{tj:scaling1} has the solution
\begin{equation}
\tj(l) = \alpha \tan\biggl[ \acot\alpha + \frac{\alpha \rho_0 J_z}{r}
  \bigl(1 - e^{-rl}\bigr) \biggr] .
\end{equation}

For antiferromagnetic bare exchange ($J_z>0$), the Kondo phase occupies the
region of parameter space in which there is a solution $0\le l_K<\infty$ of the
equation $\tj(l_K)=\infty$, i.e.,
\begin{equation}
\acot\alpha + \frac{\alpha \rho_0 J_z}{r} > \frac{\pi}{2}
\end{equation}
Thus, the Kondo phase extends over $J_z>J_{z,c}(\alpha)$ where
\begin{equation}
\rho_0 J_{z,c}(\alpha) = r \, \frac{\atan\alpha}{\alpha}.
\end{equation}
For $\alpha\ll 1$ (weak anisotropy),
\begin{equation}
\rho_0 J_{z,c} \simeq r \bigl(1 - {\ts\frac{1}{3}}\alpha^2\bigr),
\end{equation}
which reduces for $\alpha\to 0$ to the standard result \cite{Withoff:90}
$\rho_0 J_{z,c}=\rho_0 J_{\!\p,c}=r$.
For $\alpha\gg 1$ (strong anisotropy),
\begin{equation}
\rho_0 J_{z,c} \simeq \frac{r\pi}{2\alpha} \biggl(1-\frac{2}{\pi\alpha}\biggl),
\end{equation}
which implies that the Kondo phase occupies the region $J_{\!\p}>J_{\!\p,c}$
where
\begin{equation}
\label{J_pc:1}
\rho_0 J_{\!\p,c} \simeq \frac{r\pi}{2} \biggl(1-\frac{2}{\pi\alpha}\biggl).
\end{equation}

For ferromagnetic bare exchange ($J_z<0$), the condition for entry into
the Kondo regime becomes $\tj(l_K)=-\infty$, which is met for some finite
$l_K$ provided that
\begin{equation}
\acot\alpha + \frac{\alpha \rho_0 J_z}{r} < -\frac{\pi}{2}.
\end{equation}
Due to the dependence of $\alpha$ on $J_z$, this inequality is more likely to
be satisfied for smaller values of $|J_z|$ than for larger values.
Therefore, the Kondo phase extends over the region $J_z>J_{z,c}(\alpha)$, where
\begin{equation}
\rho_0 J_{z,c}(\alpha) = -\frac{r}{\alpha} (\pi - \atan\alpha) .
\end{equation}
For $0<\alpha\ll 1$ (weak anisotropy),
\begin{equation}
\rho_0 J_{z,c} \simeq -\frac{r\pi}{\alpha}\biggl(1-\frac{\alpha}{\pi}\biggl),
\end{equation}
while for $\alpha\gg 1$ (strong anisotropy),
\begin{equation}
\rho_0 J_{z,c} \simeq -\frac{r\pi}{2\alpha}\biggl(1+\frac{2}{\pi\alpha}\biggl),
\end{equation}
so the Kondo phase spans $J_{\!\p}>J_{\!\p,c}(\alpha)$ where
\begin{equation}
\label{J_pc:2}
\rho_0 J_{\!\p,c}(\alpha)
  \simeq \frac{r\pi}{2} \biggl(1+\frac{2}{\pi\alpha}\biggl).
\end{equation}

\subsubsection{Easy-axis anisotropy}

For $|J_z|> J_{\!\p}>0$, $\alpha$ defined in Eq.\ \eqref{alpha:def}
satisfies $-1 < \alpha < 0$
and the solution of Eq.\ \eqref{tj:scaling1} is
\begin{equation}
\tj(l) = \alpha \coth\biggl[ \atanh\alpha - \frac{\alpha \rho_0 J_z}{r}
  \bigl(1 - e^{-rl}\bigr) \biggr] .
\end{equation}

For antiferromagnetic bare exchange ($J_z>0$), the Kondo phase spans
the region
in which
\begin{equation}
\atanh\alpha - \frac{\alpha \rho_0 J_z}{r} > 0,
\end{equation}
i.e., the region $J_z>J_{z,c}$ where
\begin{equation}
\label{J_z,c:easy-axis}
\rho_0 J_{z,c} = r \, \frac{\atanh\alpha}{\alpha} .
\end{equation}
For $|\alpha|\ll 1$ (weak anisotropy),
\begin{equation}
\rho_0 J_{z,c} \simeq r \bigl(1 + {\ts\frac{1}{3}}\alpha^2\bigr)
\end{equation}
while for $\alpha\to -1^+$ (strong anisotropy),
\begin{equation}
\rho_0 J_{z,c} \simeq \frac{r}{2} \: \ln \!\frac{2}{1+\alpha} .
\end{equation}

For $J_z<-J_{\!\p}<0$, it is straightforward to see that
$|\alpha|\le \tj(l)<1$ for all $l>0$ and the system always lies in the
LM phase.

\subsubsection{$XY$ exchange anisotropy}

In the special case $J_z=0$ of pure-$XY$ bare exchange coupling, the
scaling in Eq.\ \eqref{tj:def} can be replaced by
\begin{equation}
\label{tjp:def}
\tJ_z = \tjp(l) \, J_{\!\p} \, e^{-rl} ,
\end{equation}
which converts Eq.\ \eqref{J_z:ode-scaling} to
\begin{equation}
\label{tjp:scaling}
\frac{d\tjp}{dl}
  = \bigl( \tjp^2 + 1 \bigr) \rho_0 J_{\!\p} \, e^{-rl}
\end{equation}
with initial condition $\tjp(0)=0$. The equation has solution
\begin{equation}
\tjp(l) = \tan\biggl[ \frac{\rho_0 J_{\!\p}}{r} \bigl(1-e^{-rl}\bigr)\biggr].
\end{equation}
In the Kondo phase, there must be an $l_K$ ($0<l_K<\infty$) such that
$\tjp(l_K)=\infty$, a condition that is satisfied for $J_{\!\p}>J_{\!\p,c}$
where
\begin{equation}
\rho_0 J_{\!\p,c} = \frac{r\pi}{2}.
\end{equation}
As one would expect, this result coincides with the limits
$\alpha\to\infty$ of Eqs.\ \eqref{J_pc:1} and \eqref{J_pc:2}.

\subsubsection{Comparison with NRG}
\label{subsubsec:compare:NRG}

The preceding results for the location of the phase boundary as a function of
$\alpha$ and the sign of $J_z$ can be re-expressed as the statement that for
any value of $J_z$, the Kondo phase occupies the region
$J_{\!\p}>J_{\!\p,c}(J_z)$, where $J_{\!\p,c}$ is a monotonically decreasing
function of $J_z$ that has the following limiting forms:
\begin{subequations}
\label{J_pc:summary}
\begin{align}
\label{J_pc:limit1}
\rho_0 J_{\!\p,c}
  &\simeq \rho_0 |J_z| \biggl[ 1 + \frac{1}{2}
     \biggl(\frac{r\pi}{\rho_0 J_z}\biggr)^2\biggr]
  && \text{for } 1 \gg -\rho_0 J_z \gg r\pi, \\[1ex]
\label{J_pc:limit2}
\rho_0 J_{\!\p,c}
  &\simeq r\pi/2 - 2 \rho_0 J_z/\pi
  && \text{for } \rho_0 |J_z| \ll r, \\[1ex]
\label{J_pc:limit3}
\rho_0 J_{\!\p,c}
  &\simeq \ds r - \half(\rho_0 J_z - r)
  && \text{for } |\rho_0 J_z\!-\!r|\ll r/3, \\[1ex]
\label{J_pc:limit4}
\rho_0 J_{\!\p,c}
  &\simeq 2 \rho_0 J_z \exp(-\rho_0 J_z/r)
  && \text{for } r \ll \rho_0 J_z\ll 1 .
\end{align}
 \end{subequations}
In the limit $r\to 0$, these expressions reproduce the standard
result \cite{Anderson:70b} $J_{\!\p,c}=|J_z|\theta(-J_z)$. The purpose of
this section is to test these statements based on poor man's scaling
against nonperturbative NRG calculations.

\begin{figure}[t]
\centering
\includegraphics[width=\columnwidth]{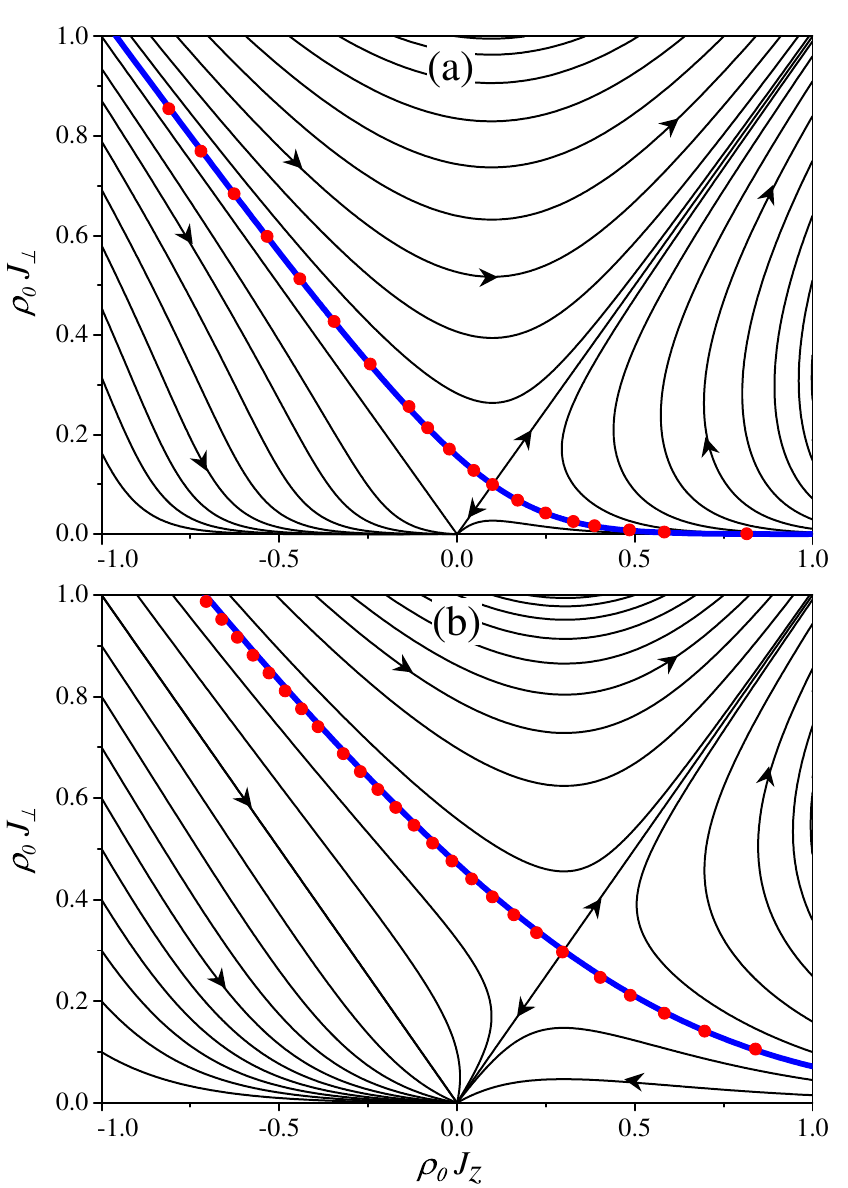}
\caption{\label{fig:flows:r>0}
Scaling trajectories for the pseudogap Kondo model with (a) $r=0.1$,
(b) $r=0.3$, calculated via numerical iteration of Eqs.\ \eqref{aniso:scaling}.
Arrows indicate the direction of flow under reduction of the half-bandwidth
$D$. Thick lines show trajectories that flow to the critical fixed point,
thereby defining the boundary between the LM and Kondo phases. Circles represent
points on the phase boundary as determined using the NRG, with all $J_z$ and
$J_{\!\p}$ values for given $r$ rescaled by the same multiplicative factor,
chosen so that the isotropic boundary point is located at
$\rho_0 J_z = \rho_0 J_{\!\p} = r$.}
\end{figure}

Scaling trajectories for the pseudogap Kondo model, calculated via numerical
iteration of Eqs.\ \eqref{aniso:scaling} with different starting parameters,
are plotted in Figs.\ \ref{fig:flows:r>0}(a) and \ref{fig:flows:r>0}(b) for
$r=0.1$ and $r=0.3$, respectively. Solid lines show trajectories that flow to
the fixed points of the model. Arrows on some of the trajectories show the
direction of flow of the couplings under reduction of the half-bandwidth $D$.
The phase boundary (thick line) separating the basins of attraction of the
LM fixed point ($\rho_0 J_z=\rho_0 J_{\!\p}=0$) and the Kondo fixed point
($\rho_0 J_z=\rho_0 J_{\!\p}=\infty$) was found by (a) reversing the flow of
Eqs.\ \eqref{aniso:scaling} and (b) choosing starting parameters very close
to the critical coupling $\rho_0 J_{z,c}=\rho_0 J_{\!\p,c}=r$ and lying on
either side of the the trajectory $J_z=J_{\!\p}$.
For comparison, NRG data for the phase boundary (circles) are shown, with
all values of $J_z$ and $J_{\!\p,c}$ rescaled by the multiplicative factor
that places the isotropic critical point at $\rho_0 J_z=\rho_0 J_{\!\p,c}=r$.
This $r$-dependent multiplicative factor is introduced to account both for a
known reduction in hybridization arising from the NRG discretization
\cite{Krishna-murthy:80,Gonzalez-Buxton:98} and for the effect of
higher-order terms omitted from the poor man's scaling equations
\eqref{aniso:scaling}, which shift the isotropic critical point from
$\rho_0 J_c = r$ to $\rho_0 J_c = f(r) \simeq r(1+r/2)$ \cite{Ingersent:96}.
Figure \ref{fig:flows:r>0} shows that poor man's scaling does an excellent
job of reproducing the shape of the phase boundary over the entire
region of couplings $\rho_0 |J_z| < 1$, $\rho_0 J_{\!\p} < 1$.

\begin{figure}
\centering
\includegraphics[width=\columnwidth]{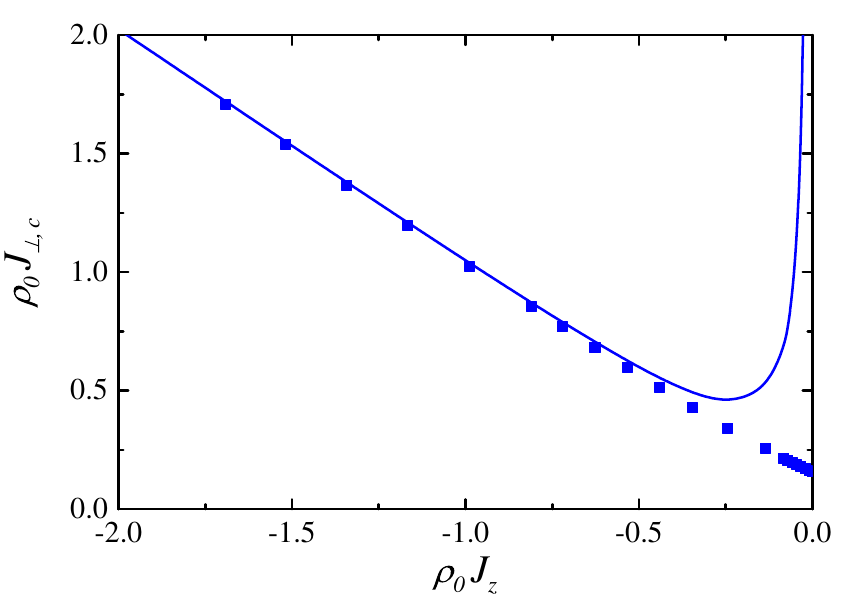}
\caption{\label{fig:limit1}
Pseudogap Kondo model phase boundary $\rho_0 J_{\!\p,c}$ vs $\rho_0 J_z$
for $r=0.1$, comparing NRG results (symbols) with the poor man's scaling
prediction for $\rho_0 J_z\ll -r\pi$ as given in Eq.\ \eqref{J_pc:summary}
(line).}
\end{figure}

\begin{figure}
\centering
\includegraphics[width=\columnwidth]{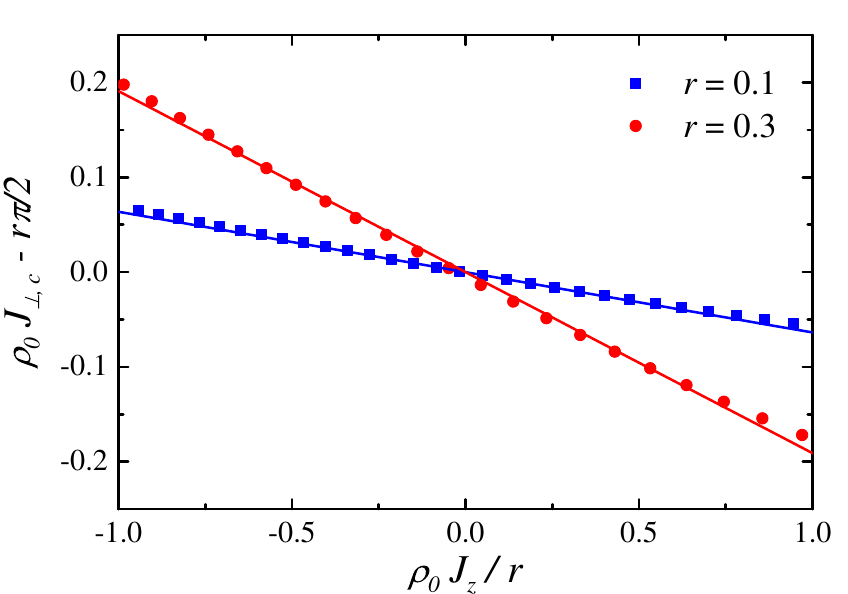}
\caption{\label{fig:limit2}
Pseudogap Kondo model phase boundary plotted as $\rho_0 J_{\!\p,c}-r\pi/2$ vs
$\rho_0 J_z/r$ for $r=0.1$ and $r=0.3$, comparing NRG results (symbols) with
the poor man's scaling prediction for $\rho_0|J_z|\ll r$ as given in Eq.\
\eqref{J_pc:limit2}
(lines).}
\end{figure}

\begin{figure}
\centering
\includegraphics[width=\columnwidth]{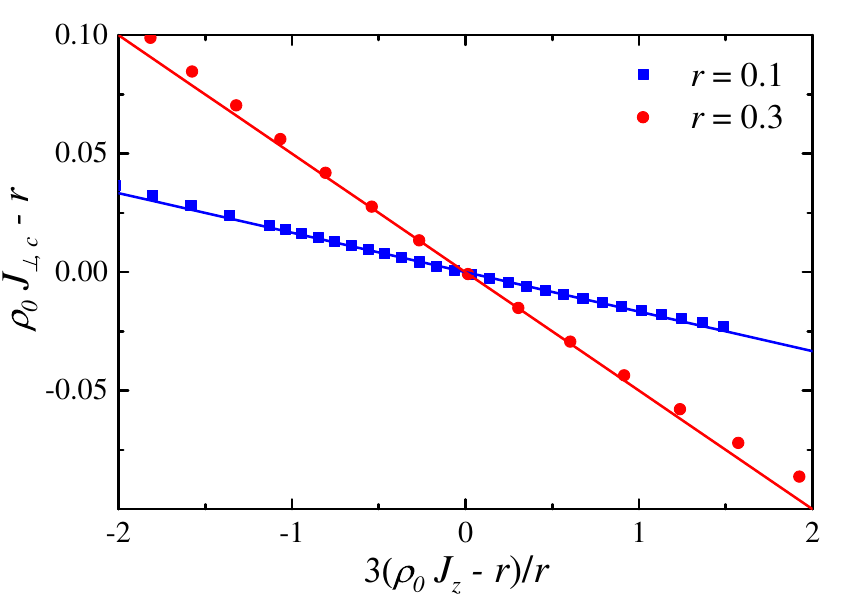}
\caption{\label{fig:limit3}
Pseudogap Kondo model phase boundary plotted as $\rho_0 J_{\!\p,c}-r$ vs
$3(\rho_0 J_z-r)/r$ for $r=0.1$ and $r=0.3$, comparing NRG results (symbols)
with the poor man's scaling prediction for $\rho_0|J_z-r|\ll r/3$ as given
in Eq.\ \eqref{J_pc:limit3} (lines).}
\end{figure}

\begin{figure}
\centering
\includegraphics[width=\columnwidth]{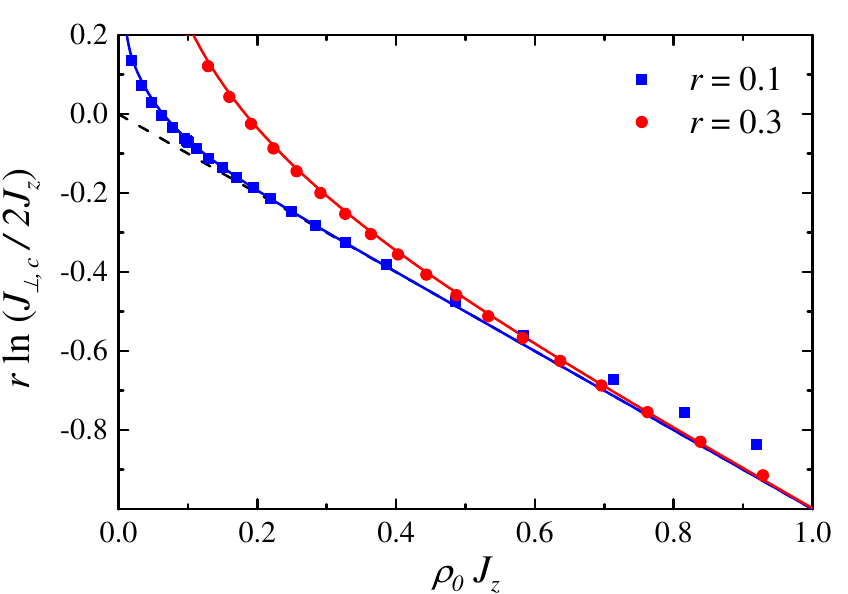}
\caption{\label{fig:limit4}
Pseudogap Kondo model phase boundary plotted as $r\ln(J_{\!\p,c}/2J_z)$ vs
$\rho_0 J_z$ for $r=0.1$ and $r=0.3$, comparing NRG results (symbols) with
the poor man's scaling prediction obtained via numerical solution of Eq.\
\eqref{J_z,c:easy-axis} (solid lines) and the asymptotic form for
$\rho_0 J_z\gg r$ from Eq.\ \eqref{J_pc:limit4} (dashed line).}
\end{figure}

A more rigorous test of the poor man's scaling is provided by Figs.\
\ref{fig:limit1}--\ref{fig:limit4}, which compare $\rho_0 J_{\!\p,c}$ vs
$\rho_0 J_z$ calculated for one of the limiting cases in Eqs.\
\eqref{J_pc:summary} (solid lines) with their NRG counterparts (symbols).
The NRG results are again scaled so that the isotropic critical point is at
$\rho_0 J_z=\rho_0 J_{\!\p,c}=r$).

Figure \ref{fig:limit1} plots the critical coupling $\rho_0 J_{\!\p,c}$ for
$r=0.1$ over a range of ferromagnetic exchange couplings $\rho_0 J_z < 0$.
Although the perturbative scaling analysis is not strictly valid for
$\rho_0 J_z \lesssim -1$, Eq.\ \eqref{J_pc:limit1}  captures surprisingly well
the variation of $\rho_0 J_{\!\p,c,\mathrm{NRG}}$ at least as far as
$\rho_0 J_z = -1.7$. For $r=0.3$, the restriction $\rho_0 J_z \ll -r \pi$
rules out the applicability of Eq.\ \eqref{J_pc:limit1} anywhere within the
range of validity of the scaling equations, so no results are shown for
this case.

In Fig.\ \ref{fig:limit2}, the critical coupling is plotted as
$\rho_0 J_{\!\p,c}-r\pi/2$ versus $\rho_0 J_z/r$ near $\rho_0 J_z=0$.
The NRG results are well reproduced by the poor man's scaling prediction Eq.\
\eqref{J_pc:limit2} over its entire range of validity $\rho_0 |J_z| \lesssim r$.
Figure\ \ref{fig:limit3} focuses on the vicinity of the isotropic critical point
at $\rho_0 J_z=\rho_0 J_{\!\p,c}=r$, plotting the critical coupling as
$\rho_0 J_{\!\p,c}-r$ vs $3(\rho_0 J_z-r)/r$. The scaling prediction in
Eq.\ \eqref{J_pc:limit3} closely reproduces the NRG results over the range
$\rho_0 |J_z-r|\ll r/3$.

Lastly, Fig.~\ref{fig:limit4} plots the critical coupling as
$r\ln (J_{\!\p,c}/2J_z)$ versus $\rho J_z$ for $0 \leq \rho_0 J_z\leq 1$.
We find that the NRG results closely follow the asymptotic form for
$r\ll \rho_0 J_z \ll 1$ [Eq.\ \eqref{J_pc:limit4}, dashed line] over
the range $0.2 \lesssim \rho_0 J_z \lesssim 0.7$ for $r=0.1$ and over
$0.6 \lesssim \rho_0 J_z \lesssim 1$ for $r=0.3$. There are minor deviations
from the asymptotic form as $\rho_0 J_z$ nears $1$ due to perturbative effects
beyond second order.
We have also plotted the poor man's scaling prediction obtained via numerical
solution of Eq.\ \eqref{J_z,c:easy-axis} (solid lines), which can be seen to
describe correctly the deviation of $J_{\!\p,c,\mathrm{NRG}}$ near $J_z=0$
from its $\rho_0 J_z \gg r$ asymptote.

The overall conclusion from Figs.\ \ref{fig:flows:r>0}--\ref{fig:limit4} is
that the poor man's scaling approach provides an excellent account of the
location of the boundary between the Kondo and local-moment phases of the
spin-anisotropic pseudogap Kondo model under conditions of strict \ph\
symmetry.

\subsection{Divergent density of states}
\label{subsec:divergent:K}

\begin{figure}
\centering
\includegraphics[width=\columnwidth]{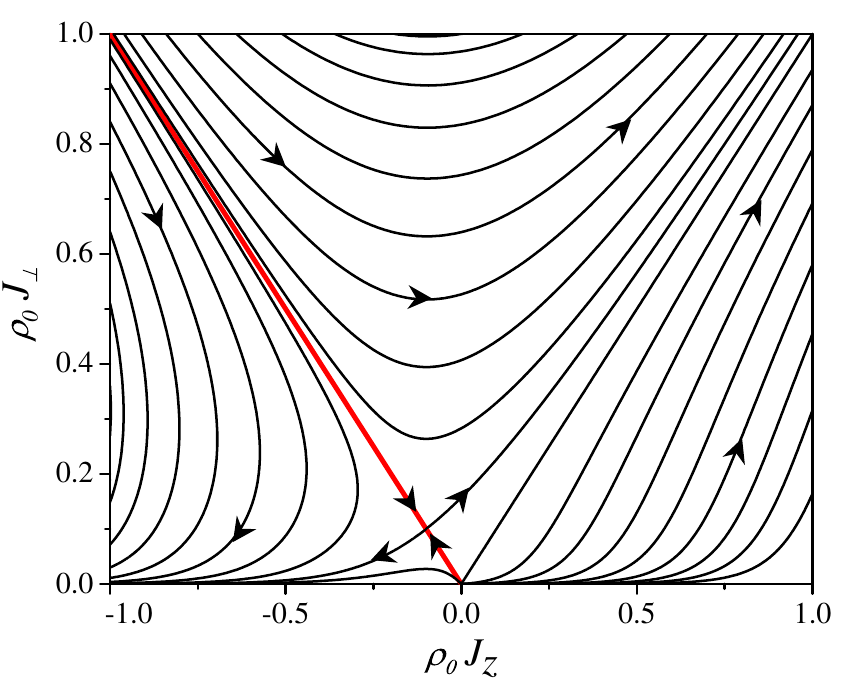}
\caption{\label{fig:flows:r=-0.1}
Scaling trajectories for the pseudogap Kondo model with $r=-0.1$,
representing a divergence of the density of states at the Fermi energy.
Arrows indicate the direction of flow under reduction of the half-bandwidth
$D$. Thick lines show trajectories that flow to the critical point,
defining the phase boundary between the ferromagnetic and Kondo phases.}
\end{figure}

Coupling a Kondo impurity to a fermionic density of states that diverges at the
Fermi level in a manner described by Eq.\ \eqref{rho:def} with $r<0$ has been
shown to yield rich physics including nontrivial quantum phase transitions
occurring for ferromagnetic exchange couplings $J<0$ \cite{Vojta:02,
Mitchell:13}. The poor man's scaling analysis of the spin-anisotropic Kondo
model applies also to cases $r<0$. Examination of Eqs.\ \eqref{aniso:scaling}
show that the poor man's scaling trajectories for $r<0$ can be obtained from
those for band exponent $|r|>0$ through the simple replacements $r\to-r$,
$\tJ_z \to -\tJ_z$. This mapping implies that the scaling trajectories for
$r<0$ should be reflections of those for $r>0$ about the axis $J_z=0$ with
reversal of the direction of flow arrows. This is illustrated in
Fig.\ \ref{fig:flows:r=-0.1}, which plots the scaling trajectories for a
representative case $r=-0.1$ over the range of exchange couplings
$-1<\rho_0 J_z<1$ and $0<\rho_0 J_{\!\p}<1$. Arrows indicate the direction of
flow of couplings with decreasing effective half-bandwidth. The model has three
stable fixed points: a ferromagnetic fixed point at
$(\rho_0 J_z,\rho_0 J_{\!\p})=(-\infty,0)$ where the impurity is locked
into a many-body spin triplet with the conduction band, the symmetric
strong-coupling fixed point at $(\rho_0 J_z,\rho_0J_{\!\p})=(\infty,\infty)$,
and an intermediate coupling fixed point at
$(\rho_0 J_z,\rho_0J_{\!\p})=(-|r|,|r|)$. The phase boundary (thick lines)
separating the ferromagnetic and strong-coupling phase is given by the condition
$J_z=-|J_{\!\p}|$, which is entirely consistent with NRG results for the
model (data not shown in Fig.\ \ref{fig:flows:r=-0.1}).

\section{Discussion}
\label{sec:discussion}

In this work, we have extended the poor man's scaling approach to analyze phase
boundaries in variants of the Anderson and Kondo impurity models in which a
power-law vanishing or divergence of the host density of states at the Fermi
energy gives rise to a nontrivial phase diagram featuring local-moment and
Kondo-screened ground states. In the regime of weak-to-moderate impurity-band
couplings where poor man's scaling is expected to be valid,
the predicted locations of the phase boundaries are generally in excellent
qualitative and good quantitative agreement with those obtained using the
numerical renormalization group (NRG). Although the NRG remains the most
reliable technique for treating power-law quantum impurity problems, the scaling
approach has the advantages that it is much more intuitive and it can clarify
algebraically the functional dependence of the critical impurity-host coupling
on other model parameters. Thus, poor man's scaling retains considerable value
even for quantum impurity problems where two or more competing RG flows give
rise to different possible infrared-stable fixed points separated by quantum
phase transitions.

Despite its successes demonstrated in Secs.\ \ref{sec:Anderson} and
\ref{sec:Kondo}, poor man's scaling has two significant limitations.
First, and more obviously, the approach is perturbative in the
impurity-band coupling and is unable to describe physics at strong coupling.
In the pseudogap Anderson model, a reliable calculation of the critical
hybridization based on poor man's scaling alone is possible for all $r>0$
only for $0<-\Ed\ll U/2$ (on the \ASCm\ side) or $0 < U+\Ed\ll U/2$ (on the
\ASCp\ side). Near the \ph-symmetric point $\Ed=-\half U$, the method breaks
down for $r\gtrsim \frac{1}{3}$. This is clear to see for $r>\half$ because
$\G_c(U,\Ed)$ diverges as $\Ed\to-\half U$ [see Fig.\
\ref{fig:PAM_phase}(b)] and therefore any phase boundary lies outside the
perturbative regime (as is also the case for the corresponding Kondo
model). For $\frac{1}{3}\lesssim r<\half$, $\G_c(U,\Ed)$ remains finite
for all $-U<\Ed<0$ [see Fig.\ \ref{fig:PAM_phase}(b)] but, as discussed in
Ref.\ \onlinecite{Gonzalez-Buxton:98} and in Sec.\ \ref{subsec:r<1} above,
the strong-coupling phases are accessed directly from mixed valence, and in
such cases we have been unable to find a scaling criterion for locating
the phase boundaries.

A second deficiency of poor man's scaling is that it does not seem to be
capable of reproducing the full RG fixed-point structure identified using
the NRG \cite{Gonzalez-Buxton:98}. Scaling Eq.\ \eqref{Ed+halfU:scaling}
and its counterpart $d\tilde{K}/d\tD=r\tilde{K}/\tD$ for the  potential
scattering in the pseudogap Kondo model both indicate that \ph\
asymmetry is an irrelevant perturbation about the symmetric plane
$\tEd=-\half\tU$. This is consistent with NRG results for band exponents
on the range $0<r\le r^*\simeq 3/8$, where a single \ph-symmetric
quantum critical point (QCP) governs the physics all over the phase
boundary between the LM and strong-coupling phases shown in Fig.\
\ref{fig:PAM_phase}(a). However, there also exists a range $r^*<r<\half$
in which the boundary between the LM phase and each strong-coupling
phase (SSC, \ASCm, and \ASCp) is governed by a different QCP. Within
this second range of band exponents, poor man's scaling cannot detect
that \ph\ asymmetry is a relevant perturbation that causes flow from the
symmetric QCP to one or the other of the two asymmetric QCPs [as illustrated
schematically for the pseudogap Kondo model in Fig.\ 16(b) of Ref.\
\onlinecite{Gonzalez-Buxton:98}].
This is a quite subtle aspect of the pseudogap Kondo and Anderson models
that even much more sophisticated RG treatments are unable to fully
capture \cite{Fritz:04}.

\acknowledgments

This work has been supported in part by NSF MWN Grants No.\ 0710540 (M.C.,
K.I.) and No.\ DMR-1107814 and No.\ DMR-1508122 (T.C., K.I.). A.M.\ was
supported by the University of Florida REU Site in Materials Physics under
NSF Grant No.\ DMR-1156737 with additional support from the US Department of
Defense. We thank E.\ Kogan for bringing Ref.\ \onlinecite{Kogan:17} to our
attention and for useful discussions.


\begin{thebibliography}{66}
\bibitem{Cheng-note}
  Reference \onlinecite{Cheng:13} states incorrectly that meeting
  the conditions for case (5) necessarily results in entry into a
  strong-coupling phase.
\bibitem{impurity-props}
  The impurity contribution to a physical property is defined in the standard
  manner (see, e.g., Ref.\ \protect\onlinecite{Krishna-murthy:80}) as
  the difference between the total value of the property for the system
  including the impurity and the total value of the same property for the
  host alone.
\end{thebibliography}
\end{document}